\documentclass[11pt,reqno]{amsart}

\pdfoutput=1

\usepackage{a4wide}

\usepackage[utf8]{inputenc} 
\usepackage{textcomp} 
\usepackage{graphicx}  
\usepackage{flafter}  
\usepackage{natbib} 

\usepackage{amsmath,amssymb}  
\usepackage{bm}  
\usepackage{latexsym} 
\usepackage{calrsfs}

\usepackage{appendix}

\allowdisplaybreaks

\usepackage{color}
\definecolor{dullmagenta}{rgb}{0.4,0,0.4}   
\definecolor{darkblue}{rgb}{0,0,0.4}

\usepackage{memhfixc}  
\usepackage{pdfsync}  

\usepackage{enumitem}

\usepackage{subfigure}

\usepackage[ruled,noend,noline,slide]{algorithm2e}

\newcommand{\E}{{\mathbb E}}
\newcommand{\Pb}{{\mathbb P}}

\theoremstyle{remark}
\newtheorem{remark}{Remark}[section]

\title{Sampling-based estimation of in-degree distribution \\ with applications to directed complex networks}

\usepackage[foot]{amsaddr}
\author{Nelson Antunes$^\ddagger$}
\address{$^\ddagger$Center for Computational and Stochastic Mathematics, University of Lisbon, Avenida Rovisco Pais 1049-001, Lisbon, Portugal.}
\email{$^\ddagger$nantunes@ualg.pt}
\author{Shankar Bhamidi$^\diamondsuit$,  Tianjian Guo$^\dagger$, Vladas Pipiras$^\S$}
\address{$^{\diamondsuit,\dagger,\S}$Department of Statistics and Operations Research, University of North Carolina, CB 3260, Chapel Hill, NC 27599, USA.}
\email{$^\diamondsuit$bhamidi@email.unc.edu}
\email{$^\dagger$tjguo97@live.unc.edu}
\email{$^\S$pipiras@email.unc.edu}
\author{Bang Wang$^\star$}
\address{$^\star$Department of Statistics, University of Pittsburgh, 1800 Wesley W. Posvar Hall, 230 S Bouquet Street, Pittsburgh, PA 15260, USA.}
\email{$^\star$baw90@pitt.edu}

\date{\today}

\begin{document}

\pagestyle{plain}

\begin{abstract}
 The focus of this work is on estimation of the in-degree distribution in directed networks from sampling network nodes or edges. A number of sampling schemes are considered, including random sampling with and without replacement, and several approaches based on random walks with possible jumps. When sampling nodes, it is assumed that only the out-edges of that node are visible, that is, the in-degree of that node is not observed. The suggested estimation of the in-degree distribution  is based on two approaches. The inversion approach exploits the relation between the original and sample in-degree distributions, and can estimate the bulk of the in-degree distribution, but not the  tail of the distribution. The tail of the in-degree distribution is estimated through an asymptotic approach, which itself has two versions: one assuming a power-law tail and the other for a tail of  general form. The two estimation approaches are examined on synthetic and real networks, with good performance results, especially striking for the asymptotic approach.
\end{abstract}

\maketitle

\section{Introduction}
\label{s:intro}

Driven by the explosion of data on a host of real world networks, the last few years have seen vigorous activity from a number of communities including computer science, statistical physics and the social sciences to develop methodology to explore large scale networks as well as formulate  network models to understand the emergence of various properties of real world systems such as the high degree of clustering, heavy tailed degree distribution and so on \cite{albert2002statistical,newman2018networks}. 

One particular corner of this vast field of great importance is the setting of \emph{directed networks}. Real world examples include:

\begin{enumerate}[label=(\emph{\roman*}),align=right]
	\item {\bf Information and social networks:} Canonical examples of these objects include the internet at the webpage level with a directed edge from page $A$ to $B$ if webpage $A$ has a link to $B$ \cite{broder:etal:2000,brin:page:1998} or Twitter networks at various levels including edges from person $A$ to $B$ if person $A$ follows $B$ or Twitter event networks over a fixed duration where one follows particular hashtags and there is an edge from $A$ to $B$ if $A$ retweets $B$ in the time interval of interest \cite{beguerisse2014interest,golder2010structural}. One need not overstate the impact of the above networks both for every day activities as well as in determining politics and world events such as the Arab Spring \cite{khondker2011role}.  
	\item {\bf Banking and financial networks:} Here examples include transnational corporations with directed (weighted) edges denoting ownership fractions \cite{vitali:2011}, world trade networks with a directed edge from country $i$ to $j$ indicating positivity of export flow from country $i$ to $j$ \cite{fagiolo:2007} or lending activity between different banks \cite{clemente:2018} to measure exposure of banks to risks. A multitude of questions have been studied ranging from exploratory analysis including clustering and statistics for ``global corporate control'' \cite{vitali:2011} of these directed networks to quantifying risk for  contagion and the cascading of losses through such interlinked systems. 
	\item {\bf Citation networks:} Here the data object consists of papers in a single discipline such as condensed matter physics or across multiple disciplines \cite{redner2004citation,vazquez2001statistics}. While the data can be parsed as both directed or undirected networks at various resolutions, one canonical object of interest is via a directed network where we put an edge from paper $A$ to $B$ if paper $A$ cites $B$. Analysis range from exploration and clustering into sub-communities (see \cite{barabasi2002evolution,newman2004coauthorship} and the references therein)  to detecting emerging areas or understanding the evolution of these networks over time.
	\item {\bf Trophic networks in ecology:} Here one is interested in feeding relationships amongst species with directed edge from $A$ to $B$ if species $A$ preys on $B$ (in many cases vertex sets can be more general than species ranging from ``taxonomically related groups of species to whole kingdoms $\ldots$ or even non-living organic matter'' \cite{pascual2005ecological}). Questions of interest include estimation of such food webs from data,  stability of the resulting systems to invasion of new species or fluctuations in density of existing species, robustness including the number of secondary extinctions triggered by loss of primary species as well as classification of keystone species. 
\end{enumerate}

\begin{figure}[t]
  \begin{center}
        \subfigure[\tiny Out-degree distribution]{\label{f:Amazon-intro-out}\includegraphics[scale=0.45]{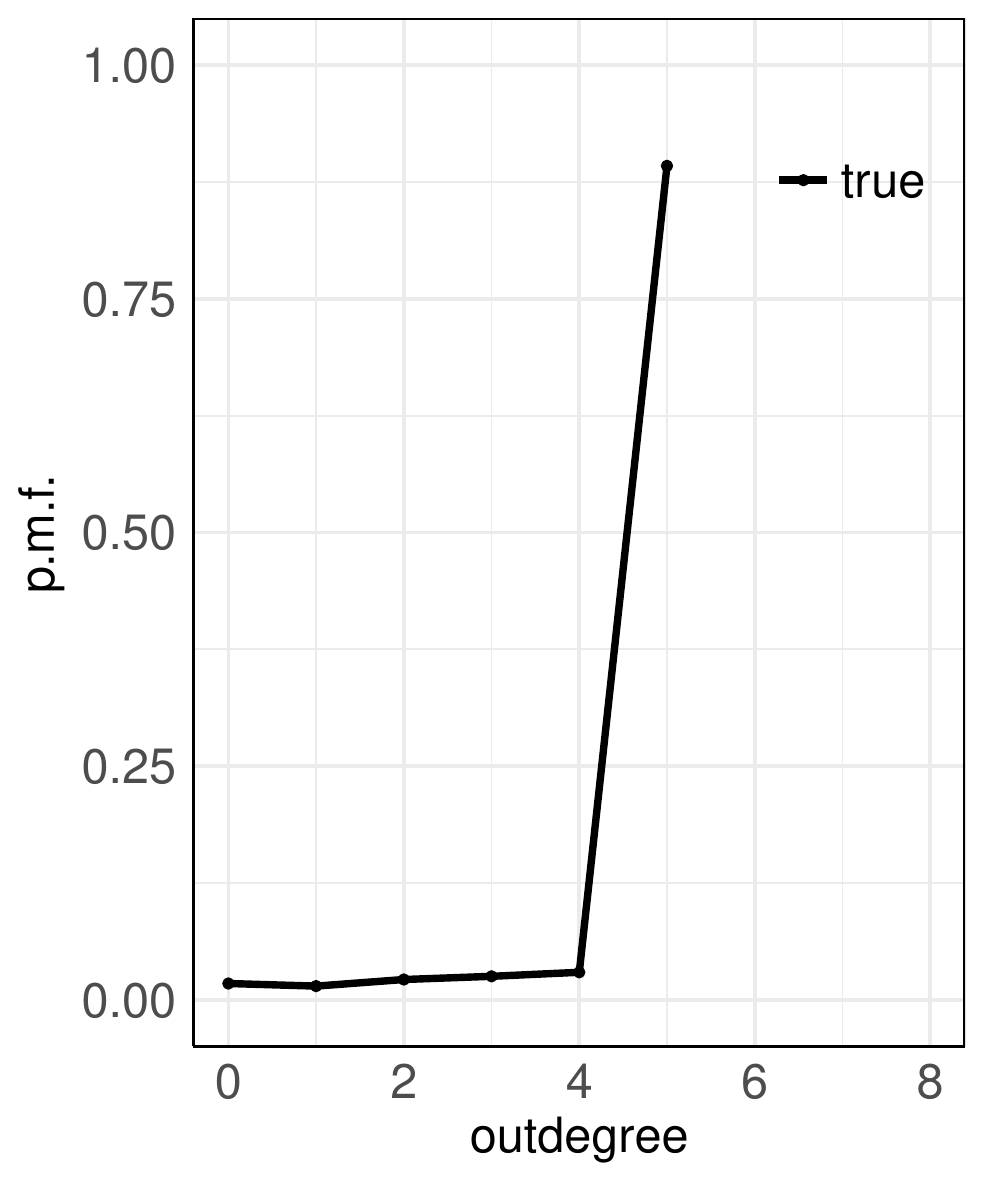}}  
        \subfigure[\tiny In-degree distribution]{\label{f:Amazon-intro-in}\includegraphics[scale=0.45]{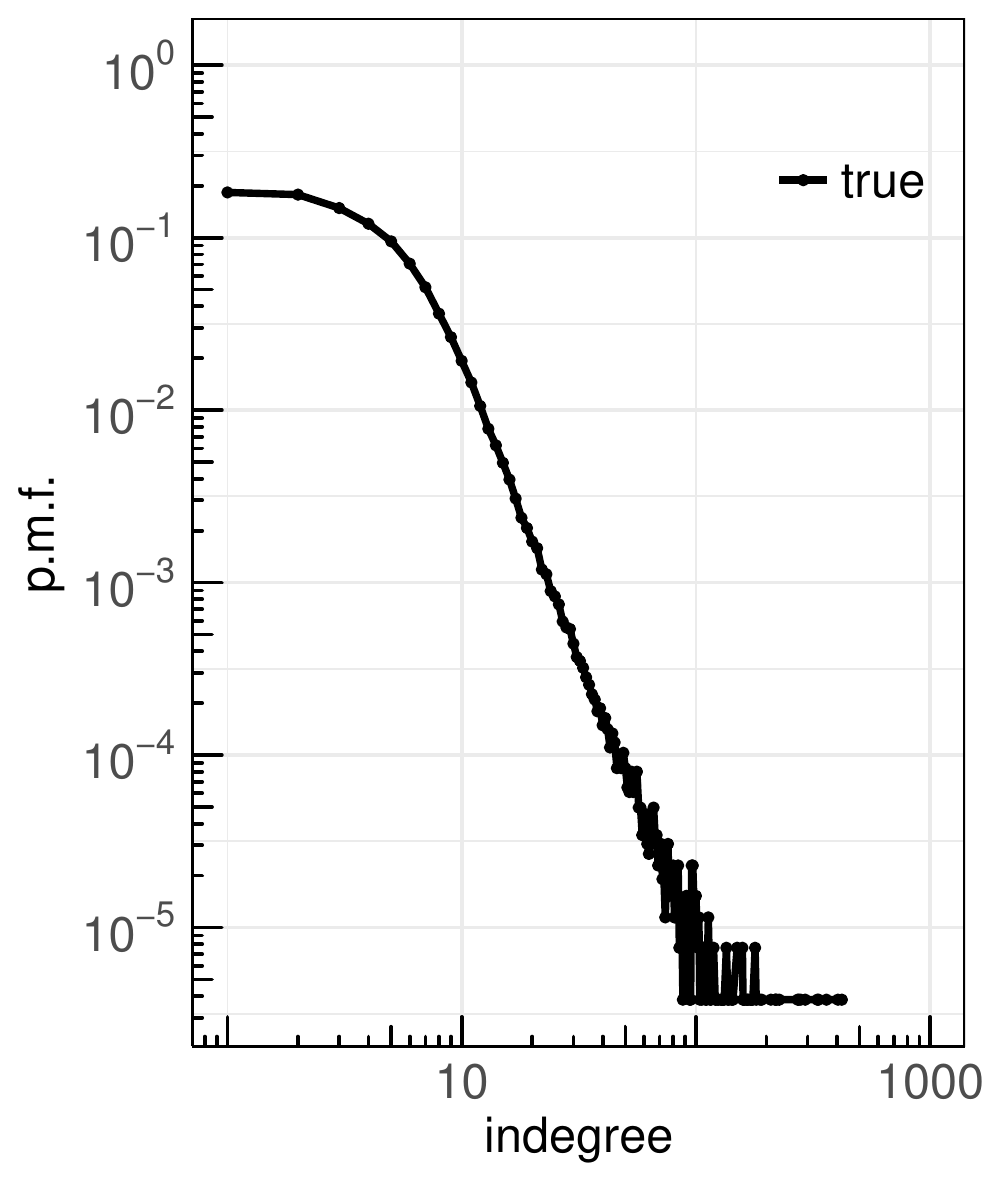}}  
        \subfigure[\tiny Inversion with RVS]{\label{f:Amazon-intro-invers}\includegraphics[scale=0.45]{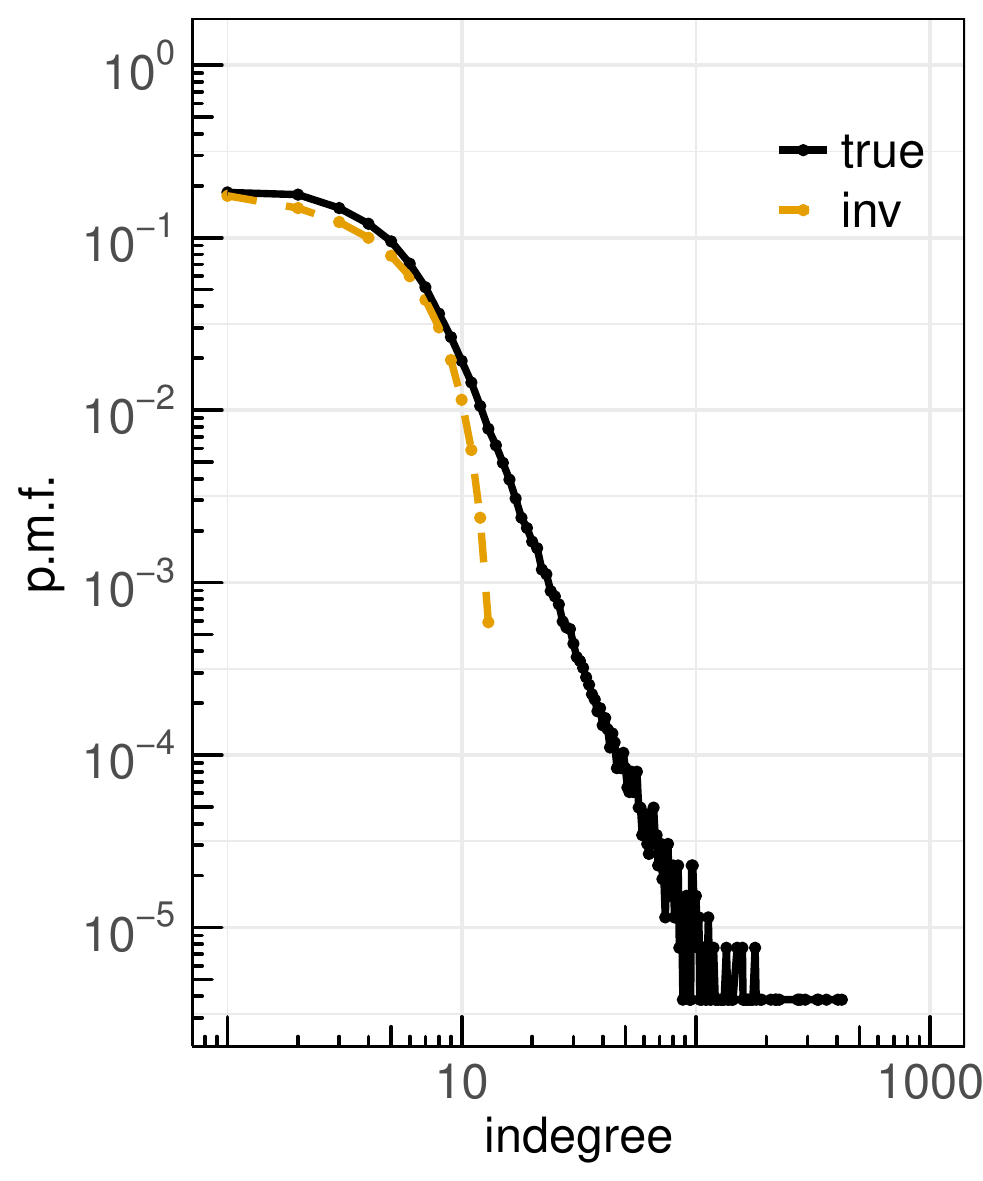}}
  \end{center}
  \caption{Amazon product co-purchasing network.}
  \label{f:Amazon-intro}
\end{figure}

For many real directed networks, the in-degrees (i.e.\ the number of in-edges) of nodes and their distribution are of particular interest. To fix ideas, consider, for example, the Amazon product co-purchasing network \cite{Leskovec:2007}, where a directed edge from product $A$ to product $B$ indicates that after buying $A$, a customer also would often purchase $B$,\footnote{There seems to be some confusion in the literature on whether to call $B$ co-purchased with $A$, or the other way around, and we will avoid the term ``co-purchased'' altogether.} after the latter appears in the Amazon recommendation feature ``Customers Who Bought This Item Also Bought This.''  The number of recommended products $B$ to purchase along with $A$ is generally small. As a result, the out-degree  of a node (product) $A$ is limited (in the considered dataset, it is limited by $5$), and the out-degree distribution is not particularly interesting, as depicted in Figure \ref{f:Amazon-intro-out}.  For this network, the in-degrees and their distribution are of greater interest. For example, the nodes with high in-degrees are the products $B$ that are also purchased when buying other products. The range and the shape of the in-degree distribution could also be of interest. The in-degree distribution for the Amazon product co-purchasing network is depicted in Figure \ref{f:Amazon-intro-in}, on the log-log scale, and the straight tail suggests, in particular, that the distribution has a power-law tail. Similar observations apply to many other directed networks, for example, citation networks that will also be considered for illustration in this work, or web link networks.

Furthermore, many real directed networks are extremely large, with large numbers of nodes and edges. Even the Amazon product co-purchasing network discussed above, which can be considered on the ``smaller'' side, already has 262,111 nodes and 1,234,877 edges. The Internet at the webpage level is estimated to have over 1.5 billion nodes (webpages). For such networks, it may be prohibitively expensive to gather all the information necessary to produce the in-degree distribution. To overcome this issue, sampling of nodes or edges seems to be a natural approach where the quantity of interest (the in-degree distribution in our case) would have to be inferred from the sampled data. Not surprisingly, sampling has been used extensively to deal with large networks and to infer their various quantities of interest  but the in-degree distribution has received less attention yet. Some related references are discussed below, after we first describe our work in some detail.

We are thus interested in inference of the in-degree distribution through sampling network nodes and edges. Several key issues arise  that we shall discuss briefly, with the goal of indicating our contributions as well:
\begin{itemize}
	\item Inherent difficulty of the problem: the latent nature of node in-degrees.
	\item Sampling schemes to use: novel schemes based on random walks.
	\item Inference approaches from statistical standpoint: application of penalized inversion approach and a new asymptotic approach.
\end{itemize}
An inherent difficulty in inferring the in-degree distribution is that, for example, when sampling a node, its in-degree is not observed but rather only its out-edges are seen, which contribute to the sample in-degrees of its neighbors. For example, this is very different from inferring the out-degree (or just the degree for undirected networks) distribution since the true out-degree of a sampled node is observed.

{\bf Sampling schemes to use: novel schemes based on random walks.} Another feature of our problem is the potential availability of various sampling schemes to use. We shall consider below random node (vertex) and random edge sampling schemes, with and without replacement, where nodes or edges are selected uniformly at random. These sampling schemes will be abbreviated as RVS and RES, respectively. 
 We shall also introduce several sampling schemes based on random walks, with occasional jumps to nodes or edges chosen at random from the network. When performing these random walks, the sample in-degree information will be collected either from sampling nodes or edges, and in their stationary regimes, the random walks will effectively sample nodes or edges as RVS or RES (with  replacement). For one of the random walks that collects information from sampled nodes, its stationary distribution will be that of RVS by employing the standard Metropolis-Hastings technique to correct for the ``bias'' in its preliminary version, whose stationary distribution will have a simple and closed form. This should be contrasted to the celebrated PageRank distribution and the corresponding random walk on a network, where the walk follows one of the out-edges of a node at random or jumps to a node chosen uniformly at random from the network. The PageRank distribution does not have a simple form and the ``bias'' cannot be removed easily, though some attempts have been made as in [16]. Further discussion and references concerning the extensive use of random walks to sample networks can be found below. We should also note that though related, the PageRank distribution is not the in-degree distribution.

{\bf Inference approaches: application of penalized inversion approach.} Turning to statistical inference, the problem of inference for the in-degree distribution from  sampled data is that of statistical inversion. More specifically, if $d=(d(j), \, j=0,\ldots,J)$, is the in-degree distribution of a network and $d_s=(d_s(j'), \, j'=0,\ldots,J')$ is its sample counterpart, we first show that
\begin{equation*}
d_s = P_s d
\end{equation*}
for a specified matrix $P_s$ which depends on the chosen sampling scheme. Part of our contributions is deriving the analytical expressions for $P_s$ under the considered sampling schemes. An inversion estimator of $d$ is defined as $\widehat d = P_s^+ \widehat d_s$, where $P_s^+$ is a suitable inverse of $P_s$ and $\widehat d_s$  refers to the  distribution of sample in-degrees. A penalized version of this ``naive'' estimator will also be considered, following the work of 
\cite{zhang:etal:2015}, in order to improve on its performance. The performance of this penalized estimator for the Amazon product co-purchasing network is illustrated in Figure \ref{f:Amazon-intro-invers}, when using RVS with the node sampling probability of $15\%$.

{\bf Inference approaches: a new asymptotic approach.} As can be seen from the latter figure, the inversion approach does not work well in the tail of the in-degree distribution due to the inherent difficulty of the problem described above. To address estimation in the tail, we also study an asymptotic approach, assuming either an arbitrary or power-law tail of the in-degree distribution. This approach essentially involves a suitable rescaling of the sample in-degree distribution. See, for example, Figures \ref{fig:RVS}--\ref{fig:non-power-law}  for illustration. The asymptotic approach for the Amazon product co-purchasing network is depicted in Figure \ref{f:Amazon}. The inversion and asymptotic approaches, when combined together, can estimate the whole in-degree distribution.

Concerning  related work, sampling-based estimation of degree distribution in {\it undirected networks} was considered by many researchers, for example, 
\cite{stumpf:2005sampling}, 
\cite{Leskovec:2006}, 
\cite{ribeiro:2010estimating,ribeiro:2012estimation}, 
\cite{Gjoka:2011}, 
\cite{kurant:2011towards}, 
\cite{lee:2012beyond}, 
\cite{zhang:etal:2015}, to name but a few. Various sampling schemes and associated inference approaches for this task were considered, most often based on different forms of random walks. Other characteristics studied under sampling in undirected networks include  global clustering coefficient (e.g.\ 
\cite{ribeiro:2010estimating}), network size (e.g.\ 
\cite{katzir:2011estimating}), number of triangles (e.g.\ 
\cite{wu:2016counting}), and others.

Turning to {\it directed networks}, 
\cite{henzinger:2000near} and 
\cite{Bar-Yossef:2008} add reversed links when sampling a directed network through random walks to avoid being ``trapped''  and in this way, build over it an undirected network, which is then used to sample nodes at random. 
\cite{ribeiro:2012} also follow this idea by introducing a Directed Unbiased Random Walk (DURW), which combines the usual random walk on a directed network with occasional jumps. See also 
\cite{Murai:2017}. One of the random walks considered in this work is similar to DURW. Sampling algorithms in directed networks based on PageRank are studied in 
\cite{Leskovec:2006}, 
 \cite{Salehi:2013}. Out-degree distribution was among network characteristics studied under some of these sampling schemes. But as noted above, inference of the in-degree distribution has seemingly not been addressed yet (though discussed in e.g.\ 
 \cite{ribeiro:2012}).

The rest of this work is organized as follows. Our sampling framework is described in Section \ref{s:sampling}. The sampling schemes based on random walks are also detailed in this section. Estimation of the in-degree distribution through statistical inversion is studied in Section \ref{s:inversion}. The section is divided into a number of parts based on the sampling scheme used. 
The  asymptotic approach is described in Section \ref{s:asymptotic}. The proposed estimation methods are examined on synthetic and real networks in Section \ref{s:data}. Conclusions can be found in Section~\ref{s:conclusions}.

\section{Sampling framework and sampling schemes}
\label{s:sampling}

\subsection{Directed graphs}

Let $G_d = (V,E_d)$ be a (simple) directed graph representing the network of interest with $N_v = |V|$ number of vertices (nodes) and $N_e = |E_d|$ number of directed edges. We denote a generic vertex as $v$, $v'$ or $u$ or, if more than one vertex is considered, as $v_i$, etc. Similarly, the notation for edges is $e$, $e'$, $e_i$, etc. The subscripts ``$v$'' and ``$e$'', on the other hand, will indicate that the corresponding quantities (e.g.\ $N_v$ and $N_e$) are associated with vertices and edges, respectively.
A directed edge from a vertex $u$ to a vertex $v$ will be denoted by $(u,v)$, that is, $(u,v)\in E_d$. We shall also denote a directed edge as $(u\to v)$,  to contrast it with edges in undirected graphs that will also be considered in some instances below. We assume that each vertex in $G_d$ has at least one (either in- or out-) edge. For notational simplicity, we shall also sometimes write $v\in G_d$ (rather than $v\in V$) and $e\in G_d$ (rather than $e\in E_d$).

\subsection{In-degree and other quantities of interest}

We focus throughout this work on the in-degree of a vertex $v$, that is,
\begin{equation}\label{e:indegree-node}
	 X (v) :=  
	 |\{u: (u,v)\in E_d\}|= \sum_{k=1}^{N_v} 1_{\{(u_k,v)\in E_d\}} 
\end{equation}
and the in-degree distribution of the graph $G_d$, that is,
\begin{equation}\label{e:indegree-dist}
	d (j) := \frac{D (j)}{\sum_{j=0}^J D (j)} :=  \frac{|\{v: X (v) =j\}|}
	{N_v} = \frac{
	 \sum_{k=1}^{N_v} 1_{\{X(v_k)=j\}}}{N_v},\quad j=0,1,\ldots,J,
\end{equation}
where $D (j)$ is the number of vertices having in-degree $j$, referred to as an in-degree count, and $J(\leq N_v -1)$ is the maximal in-degree in $G_d$. Note that $D (0)$ is non-zero if $G_d$ contains vertices with out-edges but no in-edges.



We are interested in estimating the in-degree distribution (\ref{e:indegree-dist}) through sampling. This is also equivalent to estimating the in-degree counts $D(j)$. We will sample either vertices or edges, which we shall refer to as sampling objects. When sampling a vertex, we assume importantly that {\it only} out-edges and out-neighbors of the vertex are ``visible.'' This is a common scenario encountered with real networks. The corresponding out-edges make the sampling information retained for that sampled vertex. When sampling an edge, we assume the edge to be directed and to be the sampling information retained for that sampled edge. We shall denote the (possibly repeated) sampled objects as $s_i$, $i=1,\ldots,n$, further writing $s_i=v_i$, $n=n_v$ for vertices and $s_i=e_i$, $n=n_e$ for edges. The various sampling schemes considered in this work are discussed  in Section \ref{sub:SamplingMethods}.

For a collected sample of size $n$, let 
\begin{equation}\label{e:hat-Xs}
	 \widehat X_{s}(v) = \sum_{i=1}^{n} 1_{\{s_i \ \mbox{\scriptsize has a directed edge to}\ v\}},
\end{equation}
where $s_i$, $i=1,\ldots,n$, refer to the sampled objects (i.e.\ vertices or edges), and
\begin{equation}\label{e:hat-Ds}
	\widehat D_{s}(j') = \sum_{k=1}^{N_v} 1_{\{ \widehat X_{s}(v_k) =j'\}}, \quad j'=0,1,\ldots,J'.
\end{equation}
The quantities $\widehat X_{s}(v)$ and $\widehat D_{s}(j')$ are the sample analogues of $X(v)$ in (\ref{e:indegree-node}) and $D(j')$ in ($\ref{e:indegree-dist}$), respectively, as suggested by the subscript ``$s$''. Note that $J'$ in (\ref{e:hat-Ds}) need not  be the same as $J$ and could even be larger as in sampling with replacement (discussed in this section below). The 
quantities $\widehat D_{s}(j')$'s will be modified in Sections \ref{s:inversion} and \ref{s:asymptotic} below to yield suitable estimators  $\widehat D(j)$'s of $D(j)$'s.

It will also be convenient to think about $X(v)$ in $(\ref{e:indegree-node})$ and $\widehat X_{s}(v)$ in (\ref{e:hat-Xs}) for a typical vertex $v$, namely, 
a vertex chosen (uniformly) at random. Thus, we also let $X=X(v^*)$ denote the in-degree  of a vertex $v^*$ chosen at random. Note that
\begin{equation}\label{e:X-v-rand}
\Pb (X = j ) = d(j).
\end{equation}
Additionally, let $ X_s=  \widehat X_s (v^*)$  be  the sample in-degree  of a vertex $v^*$ chosen at random from the graph  after sampling. We also set  
\begin{equation}\label{e:hat-X-v-rand}
d_s (j')  :=  \Pb ( X_s =j' ), \quad D_s(j')=N_v d_s(j').
\end{equation}
Note that
\begin{equation}\label{e:E-hat-Ds-ds}
 \frac{\E \widehat D_{s}(j')}{N_v} = \E 1_{\{\widehat X_{s}(v^*) = j'\}} = \Pb ( X_s = j') = d_s (j') = \frac{D_{s}(j')}{N_v} .
\end{equation}
Note also that we do not put a hat on $X_s$ and that $D_s(j')$ involves normalization by $N_v$ (rather than, for example, by the average number of sampled vertices).



\subsection{Sampling methods}
\label{sub:SamplingMethods}

We consider several sampling schemes. At the highest level, we shall distinguish between {\bf sampling with replacement (WR)} and  {\bf sampling without replacement (NR)}, 
the latter an acronym  for  ``No Replacement.'' In sampling without replacement,  only distinct sampled objects (vertices or edges) are included in the retained information. 
In sampling with replacement,   the same sampled objects (vertices or edges) can be included in the retained information. Samplings with and without replacement are expected to be different only when the number of sampled vertices  (edges) is relatively large  compared
to the total number of vertices (edges). This is often the case in the setting of sampling networks where the percentage of  sampled vertices (edges) have ranged in 10\%-30\% of the total population \cite{zhang:etal:2015}.

At the next level, we shall distinguish between  objects that are sampled and through what method. We shall consider: {\bf random vertex sampling (RVS)}, where vertices are sampled (uniformly) at random, with or without replacement, and {\bf random edge sampling (RES)}, where  edges are sampled (uniformly) at random, with or without replacement. We shall also sometimes write, for example, RVS-NR or RES-WR, to indicate whether NR or WR is used.

In addition, we shall also consider several  {\bf random walk sampling (RWS)} schemes to sample vertices or edges with replacement. These will be constructed so that in their 
``stationary'' regimes, they will correspond to either RVS-WR or RES-WR. In this sense, RVS and RES play a central role in our analysis -- the estimation methods developed for them in subsequent sections will be used with RWS as well. RWS schemes are discussed next. 

\subsection{Random walk samplings}
We introduce below three random walk sampling (RWS) schemes: RWS1, RWS2 and RWS3. In a ``stationary'' regime, RWS1 will correspond to RVS-WR, and RWS2 and RWS3 to RES-WR. Before presenting their constructions and to facilitate the reading, we should also indicate several of their aspects that should not be surprising. Note that without any modification, a random walk on a directed graph might be ``trapped'' at a vertex without out-edges. This is often rectified by allowing the random walk to  ``backtrack,'' that is, to walk on the directed edge backwards, thus essentially making it undirected \cite{Bar-Yossef:2008}. With our RWS schemes, we shall thus be building undirected graphs on top of the directed graph as the walk progresses. In addition, it is also common to introduce the possibility of jumps with RWS \cite{Murai:2017}. We shall allow for jumps as well and their exact role will be discussed further below. In particular, the key difference between RWS2 and RWS3 is that the walk jumps to edges  in RWS2 and to  vertices in RWS3.

\subsubsection*{\bf RWS1} We construct here a RW on a suitable graph whose stationary distribution is uniform over the graph vertices, and thus corresponds to RVS-WR. Then, a sample of vertices can be selected by using such a RW and the estimation methods for RVS discussed in Sections \ref{s:inversion} and \ref{s:asymptotic}
can be applied.


In addition to the directed graph $G_d$, our random walk constructs  and uses an undirected multigraph constructed from $G_d$, which we denote by $G^i$ at step $i$. The walk will also make occasional jumps to vertices chosen at random in the graph $G_d$, the probability (rate) of which will be associated with a parameter $w\geq 0$. The case $w=0$ corresponds to no jumps, and $w=\infty$ to jumps only. The construction of the walk is summarized in the pseudo Algorithm 1. The set $S$ gathers the  sampled, possibly repeated  vertices and $R$ the previously visited vertices.
In Step (1), the undirected graph $G^{i-1}$ is augmented to $G^i$ by including all out-edges 
from a vertex $v$ added to the sample  $S$. Step (2) selects a candidate vertex $u$, which  in Step (3)  is either selected as the next vertex $v$ to be added in the sample or the same vertex $v$ is repeated in the sample. After the algorithm is run, the information kept consists of the out-edges seen from the vertices collected in $S$.

\begin{algorithm}[t]
\SetAlgoRefName{1}
\hspace*{-.2cm}{\bf Initialization:}  Choose a vertex $v$ at random from $G_d$. Set $w$, $n_v$, $i=1$, $G^0 =\emptyset$ and $S = \{v\}$ (for the sample),  $R = \emptyset$ (for the previously visited vertices). \\
\hspace{-.2cm}{\bf Loop:} While $i< n_v$:\\
 \begin{enumerate}
\item $G^i = G^{i-1} \cup \{(v,v'): (v\to v')\in G_d,\ v\notin R\}$. 
\item Generate $U_1$ uniformly on $(0,1)$ and let $\mbox{\rm deg}_i(v)$ be the degree of the vertex $v$ in $G^i$. If $U_1<\mbox{\rm deg}_i(v)/(w+\mbox{\rm deg}_i(v))$, choose the potential next vertex $u$ at random from the neighbors of $v$ in $G^i$. Otherwise, choose $u$ at random from the vertices of $G_d$.
\item Generate $U_2$ uniformly on $(0,1)$. If $U_2<(w+\mbox{\rm deg}_i(v))/(w+\mbox{\rm deg}_i(u))$, update $v$ as $u$.  Otherwise, repeat $v$ as the next vertex.  
(If $u \notin G^i$, $\mbox{\rm deg}_i(u)$ should be read as zero.)
\item  
$R\leftarrow S$, $S \leftarrow S \cup \{u\}$. 
\item $i\leftarrow i+1$.
 \end{enumerate}
\hspace{-.2cm}{\bf Output:} $S$ (the sample of size $n_v$).
\caption{RWS1} 
\end{algorithm}

A number of further comments are in place regarding Algorithm 1. We think of $G^i$ as an undirected multigraph which as $i$ increases, becomes the undirected multigraph $G^\infty$ constructed from $G_d$ by making all directed edges undirected.
The selection of $u$ in Step (2) corresponds to the transition probabilities
\begin{equation}
P_i(v,u) = \frac{1_{\{(v,u)\in G^i\}}}{w + \mbox{\rm deg}_i(v)} + \frac{w}{w + \mbox{\rm deg}_i(v)} \frac{1}{N_v},
\end{equation}
when selecting $u$ for a given $v$. The stationary distribution of these transition probabilities (in the limit of large $i$) is not uniform on the vertices. But it could be made uniform in a standard way through a Metropolis-Hastings (MH) algorithm as in Step (3) by noting that ${w+\mbox{\rm deg}_i(v)}/{w+\mbox{\rm deg}_i(u)}$ appearing in that step is the MH acceptance function. Indeed, this follows by noting that 
\begin{align}
\frac{P_i(u,v)}{P_i(v,u)} &= \Big( \frac{1_{\{(u,v)\in G^i\}}}{w + \mbox{\rm deg}_i(u)} + \frac{w}{w + \mbox{\rm deg}_i(u)} \frac{1}{N_v} \Big) \Big/   \Big( \frac{1_{\{(v,u)\in G^i\}}}{w + \mbox{\rm deg}_i(v)} + \frac{w}{w + \mbox{\rm deg}_i(v)} \frac{1}{N_v}\Big)  \nonumber \\
&= \frac{w+\mbox{\rm deg}_i(v)}{w+\mbox{\rm deg}_i(u)}.
\end{align}
%
%
As noted above, this argument should be viewed with $i=\infty$ on $G^\infty$, that is, in the limit of large $i$.

We also note that a burn-in period could be added in Algorithm 1 where the sample starts being collected only after a certain number of steps $i$ in the algorithm have been completed.  The proposed random walk, without the MH correction, is also related to the DURW studied by 
\cite{ribeiro:2012}.

\begin{remark}
There are several reasons why jumps are used in RWS1. By varying $w$ (from $w=0$ corresponding to no jumps to $w=\infty$ associated with jumps only), their effect can be assessed on the performance of the estimation methods. For example, our inversion methods are quite insensitive to the proportion of jumps but the asymptotic approaches require a fairly large proportion of jumps. A related question is why not  use only jumps or, equivalently, RVS, if such sampling is possible. From a practical perspective, RWs without any jumps would be preferred over RVS, since this is how real networks are typically explored. RWS schemes are thus viewed as primary approaches, which are supplemented by jumps to improve performance if needed. Another potential advantage of RWs with jumps over RVS is that jumps could be imposed only to an available subset of the graph vertices or even only to vertices from the undirected graph built from the RW.
\end{remark}


We turn next to RWs where in a ``stationary''  regime, edges will be sampled at random and hence the estimation based on RES-WR can be used. We consider two RW versions, RWS2 and RWS3 below, depending on whether a jump in RW chooses an edge or a vertex at random, respectively.

\subsubsection*{\bf RWS2} 
\begin{algorithm}[t]
\SetAlgoRefName{2}
\hspace*{-.2cm}{\bf Initialization:}  Choose an edge $(v_1\to v_2)$ at random from $G_d$.  Set $w$, $n_e$, $i=1$, $G^0 =\emptyset$, $S = \{(v_1\to v_2)\}$ (for the sample), $R=\emptyset$ (for the previously visited vertices) and $jump =1$.  \\
\hspace{-.2cm}{\bf Loop:} While $i< n_e$:\\
 \begin{enumerate}
\item $G^i = G^{i-1} \cup \{(v_1,v): (v_1\to v)\in G_d, v_1\notin R\} \cup \{(v_2,v): (v_2\to v)\in G_d, v_2\notin R\}$. Edges are added as undirected but their directionality in $G_d$ is kept as an attribute. Also update $R\leftarrow R\cup \{v_1,v_2\}$.
\item If $jump =1$, choose $v$ as $v_1$ or $v_2$ at random and set $jump =0$. Otherwise, set $v=v_2$.
\item Generate $U$ uniformly on $(0,1)$. If $U<1/(1+w)$, set $(v_1\to v_2)$ as an edge selected at random from $\{(v,u) \in G^i \}$ with directionality determined by the attribute in $G^i$. Otherwise, update $(v_1\to v_2)$ as an edge selected at random from $G_d$, and set $jump=1$.
\item $S \leftarrow S \cup \{(v_1\to v_2)\}$. 
\item $i\leftarrow i+1$.
 \end{enumerate}
\hspace{-.2cm}{\bf Output:} $S$ (the sample of size $n_e$).
\caption{RWS2} 
\end{algorithm}
A pseudo algorithm for this sampling scheme is given in Algorithm 2. We again think of $G^i$ as an undirected multigraph which as $i$ increases, becomes $G^\infty$, the undirected multigraph constructed from $G_d$ by making all directed edges undirected. Several further comments on the loop steps of the algorithm follow. Note that $S$ gathers the collected sample of directed edges, whereas $R$ consists of previously ``visited'' vertices (and not edges). The variable $jump$ keeps track of whether a sampled edge was selected at random through a jump (for which $jump=1$) or as an out-edge from one of the  vertices of the previously sampled edge (for which $jump=0$).
 In Step (1), we emphasize  that edges in $G^i$ are considered undirected but they are also attributed the edge direction from $G_d$. We need to keep track of the direction in the collected sample since we are interested in the in-degree distribution and use the RES estimation that employs directional edges.  In Step\ (1), the undirected graph $G^{i-1}$ is augmented to $G^i$ by adding all out-edges from the two vertices $v_1$ and $v_2$ of the directed edge $(v_1 \rightarrow v_2)$ added to the sample. In Step (2), if the vertex $(v_1 \rightarrow v_2)$ was jumped to (thus $jump=1$), one of its vertices $v_1$ or $v_2$ is chosen at random as $v$ through which the RW will further explore the graph. On the other hand, if the vertex $(v_1 \rightarrow v_2)$ was chosen through a non-jump step of the RW ($jump=0$), this means that the RW previously visited the edge where $v_2$, was one of the vertices -- in this case, the RW will further explore the graph through the vertex $v=v_2$.


The motivation for selecting a new directed edge into the sample in Steps (3) and (4) is as follows. For large $i$, we can think of $G^i$ as the undirected multigraph $G^\infty$, defined from $G_d$ as above. The graphs $G^\infty$ and $G_d$ have the same edges, except that in $G_d$, these are made directional. The key observation then is that our algorithm samples edges at random from $G^\infty$, and hence once attributes are added, also directed edges at random from $G_d$. Indeed, to prove this key observation, it is enough to show that the stationary distribution of our RW on edges is uniform. For this, note that according to our algorithm, the transition probability for the RW on $G^\infty$ is 
\begin{align}
P((v_1,v_2),(u_1,u_2)) &= \sum_{k=0}^1 \Pb((v_1,v_2) \: \mbox{to} \: (u_1,u_2)|jump =k) \Pb(jump=k) \nonumber \\
&= P_0((v_1,v_2),(u_1,u_2)) \frac{1}{1+w} + P_1((v_1,v_2),(u_1,u_2)) \frac{w}{1+w}, 
\end{align}
with
$$
P_0((v_1,v_2),(u_1,u_2))  = \frac{1_{ \{(u_1,u_2)\in N^\infty(v_2) \} }}{\mbox{deg}_\infty(v_2)} \frac{1}{1+w} + \frac{2}{\mbox{vol}(G^\infty)} \frac{w}{1+w}
$$
and 
$$
P_1((v_1,v_2),(u_1,u_2))  = \frac{1}{2} \sum_{j=1}^2 \Bigg( \frac{1_{ \{(u_1,u_2)\in N^\infty(v_j) \} }}{\mbox{deg}_\infty(v_j)} \frac{1}{1+w} + \frac{2}{\mbox{vol}(G^\infty)} \frac{w}{1+w}  \Bigg),
$$
where $\mbox{deg}_\infty(v)$ stands for the degree of $v$ in $G^\infty$, $N^\infty(v)$ refers to the neighborhood of $v$ in $G^\infty$ (that is, the edges connected to $v$), and $\mbox{vol}(G^\infty) = \sum_{v\in G^\infty} \mbox{deg}_\infty(v)$ is also twice the number of edges in the multigraph $G^\infty$. With some abuse of the notation and to keep notation simple, if there are two edges in $G^\infty$ between two vertices $u_1$ and $u_2$, we denote both of these edges as $(u_1,u_2)$. The uniform distribution $\pi((v_1,v_2)) = 2/\mbox{vol}(G^\infty)$ on the edges is the stationary distribution for the RW since $\sum_{(v_1,v_2)} P((v_1,v_2),(u_1,u_2))=1$ (i.e.\ $P$   is a doubly stochastic transition matrix). This follows from 
$\sum_{(v_1,v_2)} P_k((v_1,v_2),(u_1,u_2))=1$ for $k=0,1$ since, for example, for $k=1$, 
\begin{align}
\sum_{(v_1,v_2)} P_1 & ((v_1,v_2),(u_1,u_2)) = \sum_{(v_1,v_2)} \frac{1}{2} \sum_{j=1}^2 \Bigg ( \frac{1_{ \{(u_1,u_2)\in N^\infty(v_j) \} }}{\mbox{deg}_\infty(v_j)} \frac{1}{1+w} + \frac{2}{\mbox{vol}(G^\infty)} \frac{w}{1+w}  \Bigg) \nonumber \\
&=  \Bigg( \sum_{(v_1,v_2) = (u_1,u_2)} \frac{1}{2} \sum_{j=1}^2  \frac{1}{\mbox{deg}_\infty(v_j)}  + \sum_{v_1=u_1,(v_1,v_2)\neq (u_1,u_2)} \frac{1}{2\mbox{deg}_\infty(v_1)}  \nonumber \\
& \hspace{5cm} + \sum_{v_2=u_2,(v_1,v_2)\neq (u_1,u_2)} \frac{1}{2\mbox{deg}_\infty(v_2)}  \Bigg)  \frac{1}{1+w} +  \frac{w}{1+w}   \nonumber \\
&= \Bigg( \frac{1}{2} \sum_{j=1}^2  \frac{1}{\mbox{deg}_\infty(u_j)} + \frac{\mbox{deg}_\infty(u_1)-1}{2\mbox{deg}_\infty(u_1)} + \frac{\mbox{deg}_\infty(u_2)-1}{2\mbox{deg}_\infty(u_2)} \Bigg)  \frac{1}{1+w} +  \frac{w}{1+w} \nonumber \\
&= \frac{1}{1+w} +  \frac{w}{1+w}  = 1.
\end{align}

\subsubsection*{\bf RWS3} 
\begin{algorithm}[t]
\SetAlgoRefName{3}
\hspace*{-.2cm}{\bf Initialization:}  Choose a vertex $v$ at random from $G_d$. Set $w$, $n_e$, $i=1$, $G^0 =\emptyset$, $S = \emptyset$ (for the sample) and $R = \emptyset$ (for  the vertices visited by RW).  \\
\hspace{-.2cm}{\bf Loop:} While $|S|< n_e$:\\
 \begin{enumerate}
\item $G^i = G^{i-1} \cup \{(v,u): v\notin R,\ (v\to u)\in G_d\}$. Edges are added as undirected but their directionality in $G_d$ is kept as an attribute.
\item $R \leftarrow R \cup \{v\}$. 
\item Generate $U$ uniformly on $(0,1)$ and let $\mbox{\rm deg}_i(v)$ be the degree of the vertex $v$ in $G^i$. If $U<\mbox{\rm deg}_i(v)/(\mbox{\rm deg}_i(v)+w)$, choose an edge $(v_1\to v_2)$ at random from $\{(v,u) \in G^i\}$ with directionality determined by the attribute in $G^i$. Update $S\leftarrow S\cup \{(v_1\to v_2)\}$ and $v$ by $v_2$. Otherwise, update $v$ by  a vertex selected at random from $G_d$, without updating $S$. 
\item $i\leftarrow i+1$.
 \end{enumerate}
\hspace{-.2cm}{\bf Output:} $S$ (the sample of size $n_e$).
\caption{RWS3} 
\end{algorithm}
We finally turn to a random walk for sampling edges where jumps will be to random vertices rather than edges. Jumping to random vertices might be preferred within the architecture of many real networks. A pseudo algorithm for RWS3 is included as Algorithm 3, and we comment on its steps and motivation next.

As in RWS2, for large $i$, we think of $G^i$ as the undirected multigraph $G^\infty$ constructed from $G_d$ by making its directed edges undirected. Let also $\mbox{\rm deg}_\infty(v)$ be the degree of a vertex $v$ in $G^\infty$. When Algorithm 3 acts on $G^\infty$, we shall argue that it selects the edges of $G^\infty$ at random (and hence also the edges of $G_d$ at random, with the directionality kept as an attribute in Algorithm 3). To make such an argument, we add an ``imaginary'' vertex $g$ to $G^\infty$ and connect it to each vertex of $G^\infty$ through an edge that carries a weight $w\geq 0$. Other edges in $G^\infty$ are then assumed to have weight $1$. The resulting multigraph with the added ``imaginary'' vertex and edges is denoted $\widetilde G^\infty$. The edge weight, $w$ or $1$, plays a role only when sampling randomly from a collection of edges, in which case their weights determine the sampling probabilities. For example, two edges with weights $1$ and $w=2$ would be sampled with the respective probabilities $1/(1+w) = 1/3$ and $w/(1+w) = 2/3$.

Now, note that Step (3) of Algorithm 3 can be reformulated for $\widetilde G^\infty$ by stating that an edge is selected randomly from $\{(v,u)\in \widetilde G^\infty\}$, with the edge weights determining sampling probabilities. Indeed, for such sampling, an edge from $G^\infty$ would be selected with the probability $\mbox{\rm deg}_\infty(v)/(\mbox{\rm deg}_\infty(v)+w)$, and the added edge to the ``imaginary'' vertex $g$ would be selected with the probability $w/(\mbox{\rm deg}_\infty(v)+w)$. In the latter case, note that the next selection would be an edge selected at random (with equal probabilities) from the ``imaginary'' edge back to a random vertex in $G_d$. This is now captured in Step (3) of Algorithm 3 through a uniform random variable $U$. 

Let $\widetilde S$ be the sample of edges collected through this reformulated Algorithm 3 acting on $\widetilde G^\infty$, that is, $\widetilde S$ is the sample $S$ supplemented by the sampled edges connecting to the ``imaginary'' vertex. It is then enough to argue that $\widetilde S$ consists of edges selected at random from $\widetilde G^\infty$ since in that case, the edges of its subset $S$ would also be selected at random but now from $G^\infty$. What needs to be shown is somewhat trivial given that edges are selected from $\widetilde G^\infty$ by traveling from a vertex to one of its neighbors chosen according to a weighted degree distribution. Indeed, this simply follows by observing that such a RW on edges has its transition probabilities as
 \begin{equation}
P ((v_1,v_2),(u_1,u_2)) = \frac{ \mbox{\rm weight of}\ (u_1,u_2)}{\widetilde{\mbox{\rm deg}}_\infty(v_2)} 1_{\{(u_1,u_2)\in \widetilde N^\infty(v_2)\}}
 \end{equation}
and one trivially has $\sum_{(v_1,v_2)} P((v_1,v_2),(u_1,u_2)) =1$, where $\widetilde N^\infty(v)$ refers to the neighborhood of $v$ in $\widetilde G^\infty$ and 
$$
\widetilde{\mbox{\rm deg}}_\infty(v) = \sum_{(v,v')\in \widetilde N^\infty(v)} (\mbox{\rm weight of}\ (v,v'))
$$ 
to a weighted degree of $v\in \widetilde G^\infty$.


\section{Estimation by inversion}
\label{s:inversion}

We consider inference of the in-degree distribution separately for sampling without replacement (Section \ref{s:inversion-woR}) and sampling with replacement (Section \ref{s:inversion-wR}). All the proposed estimators will involve a matrix inversion in one of their key construction steps, hence the reference to estimation by inversion. Inversion can and usually should also be supplemented by a suitable weighted least-squares approach with
penalization, as explained in Section \ref{s:inversion-pen} below.

We shall describe the estimators in general terms first. They will involve coefficients specific to a sampling scheme. These coefficients will then be specified in Sections \ref{s:inversion-woR}--\ref{s:inversion-wR},  for the sampling schemes listed in Section \ref{s:sampling}, that is, RVS, RES and RWS.
The estimator $\widehat D_{s}(j)$ in (\ref{e:hat-Ds}) is  biased for the in-degree count $D(j)$, as well as $ X_{s}(v)$  in (\ref{e:hat-Xs}) for $X(v)$ (for the latter, see also Remarks \ref{r:NR-connection-HT} and \ref{r:WR-connection-HH} below). In fact, for the considered sampling schemes, 
by using (\ref{e:E-hat-Ds-ds}), we  have  that
\begin{align}
D_{s} (j') = \E \widehat D_{s} (j')= N_v \Pb ( X_s = j')    &=  N_v \sum_{j=1}^J \Pb( X_s = j' | X=j) \Pb(X=j)  \nonumber \\
&= \sum_{j=1}^J \Pb( X_s = j' | X=j)  D(j), \quad j'=0,\ldots, J', \label{e:E-hat-Ds}
\end{align}
and in a matrix form
\begin{equation}
\E \widehat D_{s} =  P_s D,   \label{e:E-hat-Ds-matrix}
\end{equation} 
where $\widehat D_{s}= (\widehat D_{s}(0),\ldots,\widehat D_{s}(J'))^T$,   $ D= ( D(0),\ldots, D(J))^T$ and  $P_s=(P_s(j',j))$ is a $(J'+1)\times (J+1)$  matrix  specific to a sampling scheme, with $P_s(j',j) = \Pb ( X_s=j'|X=j)$ in view of (\ref{e:E-hat-Ds}). The latter probability should be read as: assuming that a vertex has $j$ in-edges, what are the chances that $j'$ of these $j$ in-edges will be selected through sampling?
%

In general, $P_s$ is not a square matrix. This suggests that an unbiased estimator of the in-degree counts $D$ should be defined as 
\begin{equation}\label{e:Dj-est-NR}
	{\widehat D} = P_s^{+}  \widehat D_{s},
\end{equation}
where $\widehat D=(\widehat D(0),\ldots,\widehat D(J))^T$ and $P_{s}^{+} = (P_{s}^T P_{s})^{-1} P_{s}^T$ is the  (left) generalized inverse of $P_{s}$, provided that $P_{s}$ has full column rank. 

%
%
%
%
%
%

\subsection{Sampling without replacement}
\label{s:inversion-woR}

The matrices $P_{s}$ for the RVS and RES schemes are specified below. The section concludes with several remarks shedding further light on the estimator (\ref{e:Dj-est-NR}).

\subsubsection*{\bf RVS} For random vertex sampling without replacement,  the maximum observable in-degree is  
%
%
%
$J'=\min(J,n_v)$  and the entries of the matrix $P_s$ are

\begin{equation}\label{e:P-NR-RNS}
	P_{s}(j',j)  = 
	\left\{
	\begin{array}{cl}
		\frac{\binom{j}{j'} \binom{N_v-j}{n_v-j'}} { \binom{N_v}{n_v}},  &  \mbox{if} \:\: j=0,\ldots, J, \: j'= \max\{0, n_v - N_v +j  \}, \ldots, \min\{j,n_v\},   \\
		0 , & \mbox{otherwise}.
	\end{array}
	\right.
\end{equation}
Indeed, with the interpretation of $P_{s}(j',j)$ around (\ref{e:E-hat-Ds})--(\ref{e:E-hat-Ds-matrix}),  for a vertex to have in-degree $j'$ after sampling, it has to have in-degree $j\geq j'$ in the original graph, and then $j'$ of its $j$ in-neighbors have to be sampled and $n_v-j'$ vertices have to be sampled from the remaining $N_v-j$ vertices of the graph. 
The upper bound  $\min\{j,n_v\}$ on $j'$ is used  because the maximum observed value of $j'$  is the smaller of the sample  size $n_v$ and the value of
the in-degree $j$. Also, if $n_v >  N_v -j$,  at least  in-degree $n_v + j - N_v$ is observed. In practice, one typically has  $J< n_v$,
and therefore $P_s$  is a square matrix of dimension $(J+1) \times (J+1)$. Note also that the  columns  of the matrix $P_{s}$ correspond 
to the hypergeometric distribution with  parameters $j$, $n_v$ and $N_v$.

\subsubsection*{\bf RES} As above, we have 
$J'=\min(J,n_e)$ 
and  $P_s$ has entries
\begin{equation}\label{e:P-NR-RES}
	P_{s}(j',j)  = 
	\left\{
	\begin{array}{cl}
		\frac{\binom{j}{j'} \binom{N_e-j}{n_e-j'}} { \binom{N_e}{n_e}},  & \mbox{if} \:\: j=0,\ldots, J, \: j'= \max\{0, n_e - N_e +j  \}, \ldots, \min\{j,n_e\},   \\
		0, & \mbox{otherwise}.
	\end{array}
	\right.
\end{equation}


We note that the counts $\widehat D_{s}(j')$ defined in (\ref{e:hat-Ds}) and entering (\ref{e:Dj-est-NR}) are just the in-degree counts of the sampled graph, that is, the graph obtained by connecting the graph vertices by the directed edges retained in the used sampling procedure.

\begin{remark}\label{r:NR-upper-triangular}
As noted above, the matrix $P_{s}$ in (\ref{e:P-NR-RNS}) or (\ref{e:P-NR-RES}) is usually a square  $(J+1)\times(J+1)$ 
 matrix when sampling real networks. It  is then upper triangular, and so is its inverse $P_{s}^{-1}$ used in the definition of the estimator (\ref{e:Dj-est-NR}). In particular,  $\widehat D (j)$ will be zero for those $j$ such that $\widehat D_{s}(j')=0$ for $j'\geq j$. This  suggests that (\ref{e:Dj-est-NR}) should be used with $J$ replaced by the largest in-degree obtained after sampling, which does not require the a priori knowledge of $J$. But note that a priori knowledge (or an estimate) of $N_v$ and $N_e$ would be required to compute (\ref{e:P-NR-RNS}) and (\ref{e:P-NR-RES}).

We show in  Appendix  \ref{s:appendix} that the inverse of a square matrix $P_{s}$ in e.g.\ (\ref{e:P-NR-RNS}) for RVS  can be expressed for general $J$ as

\begin{equation}\label{e:P-NR-RNS-inverse}
	P_{s}^{-1}(j',j)  = 
	\left\{
	\begin{array}{cl}
		(-1)^{j'+j} \frac{\binom{N_v-n_v+j-j'-1}{j-j'} \binom{N_v}{j'}} { \binom{n_v}{j}},  & \mbox{if} \:\: 0 \leq j  \leq J , \  0 \leq  j' \leq j ,  \\
		0 , & \mbox{otherwise}.
	\end{array}
	\right.
\end{equation}
A similar expression can be written for RES by replacing $N_v$ and $n_v$ by $N_e$ and $n_e$, respectively. When $J$ is large, the numerical inversion of $P_s$ actually differs slightly from the theoretical inverse of $P_s$ given in (\ref{e:P-NR-RNS-inverse}), and the resulting estimators (\ref{e:Dj-est-NR}) are also numerically slightly different. 

\end{remark}

\begin{remark}\label{r:NR-connection-HT}
It is interesting to contrast the estimators defined above to the so-called Horvitz-Thompson (HT) estimation often employed in sampling without replacement (see e.g.\ 
\cite{thompson:2012}, 
\cite{tille:2006}). We noted above that the estimator $\widehat X_{s}(v)$ in (\ref{e:hat-Xs})  is biased for $X(v)$. In fact, an unbiased estimator is the HT estimator defined as 
\begin{equation}\label{e:Cv-est-HT}
	\widehat X_{\scriptsize \rm HT}(v) = \sum_{i=1}^{n} \frac{1_{\{s_i \ \mbox{\scriptsize has out-edge to}\ v\}}}{\pi(s_i)},
\end{equation}
where $\pi (s_i)$ is the probability that object $s_i$ (i.e.\ vertex  or edge) is included in the sampling procedure.
(As there are   $\binom{N_v -1}{n_v- 1}$  samples that may be chosen to include a given vertex $v$ without replacement, it follows that 
$\pi(v)=\binom{N_v -1}{n_v-1} / \binom{N_v }{n_v} = n_v / N_v$ is the probability that the vertex  is included in the sample; similarly, we have
  $\pi (e) = n_e/N_e$.)
 It then seems reasonable to set
\begin{equation}\label{e:Dj-est-HT-0}
	\widehat D_{{\scriptsize \rm HT}}(j') = \sum_{k=1}^{N_v} 1_{\{ \lfloor \widehat X_{\scriptsize \rm HT}(v_k)\rfloor  =j'\}}
\end{equation}
as an estimator for $D(j')$, where $\lfloor \cdot\rfloor$ denotes the ``floor'' integer part. But the estimator $\widehat D_{{\scriptsize \rm HT}}(j')$ will be biased for $D(j')$ as well. In fact, if developed, the whole inference procedure would essentially be the same as that described above in the section. Assuming $\pi(s_i) \equiv \pi$, note that 
$$
\widehat D_{{\scriptsize \rm HT}}(j') = \widehat D_{s}( \lfloor \pi j' \rfloor), 
$$ 
that is, the vector of $\widehat D_{{\scriptsize \rm HT}}(j')$ would consist just of the $(1/\pi)$ times repeated values of the vector of $\widehat D_{s}(j)$. One could write a formula for the vector of $\widehat D_{{\scriptsize \rm HT}}(j')$ analogous to (\ref{e:E-hat-Ds-matrix}) but the corresponding probability matrix would similarly repeat the rows of the matrix $P_{s}$. To make an inverse, the repeated rows of the probability matrix (and the corresponding repeated entries of $\widehat D_{{\scriptsize \rm HT}}(j')$) would have to be removed and the resulting inversion approach would be no different from that for $\widehat D_{s}(j)$ presented above.
\end{remark}

\subsection{Sampling with replacement}
\label{s:inversion-wR}

We now turn to sampling schemes with replacement. 
The matrices $P_{s}$ for  RNS, RES and RWS are specified below. 

\subsubsection*{\bf RVS} For random vertex sampling with replacement, the maximum in-degree is  $J'=n_v$ and  the $(n_v+1)\times (J+1)$ matrix  $P_s$ has the entries


\begin{equation}\label{e:P-WR-RNS}
	P_{s}(j',j) = 
	\left\{
	\begin{array}{cl}
		{n_v\choose j'} \Big( \frac{j}{N_v}\Big)^{j'} \Big( \frac{N_v-j}{N_v}\Big)^{n_v - j'}, &  \mbox{if} \ j\geq 1,\  j'\geq 0  \: \mbox{or}  \: j=0, \  j' \geq 1 , \\
		1, & \mbox{if}\ j=0,\ j'=0 .
	\end{array} 
	\right.
\end{equation} 
Indeed, with the interpretation of $P_{s}(j',j)$ around (\ref{e:E-hat-Ds})--(\ref{e:E-hat-Ds-matrix}),  note that 
 ${n_v\choose j'}$ selects $j'$ of the $n_v$ sampled vertices that point to a vertex with in-degree $j$, with $j/{N_v}$ being  the probability of a sampled vertex pointing to the vertex with in-degree $j$ and $(N_v-j)/{N_v}$ being the probability of the complementary event. Note  that the  columns  of the matrix $P_{s}$ correspond 
to the binomial distribution with  parameters $n_v$ and $j/N_v$.

%

\subsubsection*{\bf RES} As above,  $J'=n_e$ and  $P_s$ is a $(n_e+1)\times (J+1)$ matrix with the entries
\begin{equation}\label{e:P-WR-RES}
	P_{s}(j',j) = 
	\left\{
	\begin{array}{cl}
		{n_e\choose j'} \Big( \frac{j}{N_e}\Big)^{j'} \Big( \frac{N_e-j}{N_e}\Big)^{n_e - j'} ,& \mbox{if} \ j\geq 1,\  j'\geq 0  \: \mbox{or}  \: j=0, \  j' \geq 1,  \\
		1 , & \mbox{if}\ j=0,\ j' =0 .
	\end{array} 
	\right.
\end{equation}

We note that the counts $\widehat D_{s}(j')$ defined in (\ref{e:hat-Ds}) and entering (\ref{e:Dj-est-NR}) can be thought here as the in-degree counts  of the sampled multigraph, that is, the multigraph obtained by connecting the graph vertices by the directed and possibly repeated edges retained in the used sampling procedure.  

\begin{remark}\label{r:WR-not-upper-triangular}
In contrast to Remark \ref{r:NR-upper-triangular}, we do not have  an explicit form of the inverse of the matrix $P_{s}$ (assuming it is square).  Additionally,  note that the choice of $J$ in (\ref{e:Dj-est-NR}) is somewhat arbitrary. In practice, we take $J$ to be the largest in-degree observed in the sample. We also note that a priori knowledge (or an estimate) of $N_v$ and $N_e$ would be required to compute (\ref{e:P-WR-RNS}) and (\ref{e:P-WR-RES}). \end{remark}

\begin{remark}\label{r:WR-connection-HH}
It is interesting to contrast the estimators defined above to the so-called Hansen-Hurwitz (HH) estimation often employed in sampling with replacement (see e.g.\ 
\cite{tille:2006}, 
\cite{thompson:2012}). For example, an unbiased estimator for $X(v)$ is the HH estimator defined as 
\begin{equation}\label{e:Cv-est-HH}
	\widehat X_{\scriptsize \rm HH}(v) = \frac{1}{n} \sum_{i=1}^n \frac{1_{\{s_i\ \mbox{\scriptsize has out-edge to}\ v\}}}{p(s_i)},
\end{equation}
where $p(s_i)$ to the probability of sampling $s_i$  at step $i$ (i.e.\ a vertex $v_i$ is sampled with probability $1/N_v$ and an edge $e_i$ with probability $1/N_e$) . 
One could then set
\begin{equation}\label{e:Dj-est-HH-0}
	\widehat D_{{\scriptsize \rm HH}}(j') = \sum_{k=1}^{N_v} 1_{\{\lfloor\widehat X_{\scriptsize \rm HH}(v_k)\rfloor =j'\}}
\end{equation}
as an estimator for $D(j')$. The estimator $\widehat D_{{\scriptsize \rm HH}}(j')$ will be biased. In fact, the whole inference procedure would essentially be the same as that described above in the section. Assuming $p(s_i) \equiv p$, note that 
$$
\widehat D_{{\scriptsize \rm HH}}(j') = \widehat D_{s}(\lfloor pj' \rfloor). 
$$ 
Then, the situation is analogous to that discussed in Remark \ref{r:NR-connection-HT}, and similar conclusions can be drawn.
\end{remark}

\begin{remark}
Our distinction between samplings with and without replacement might appear somewhat restrictive in the following sense.  Note that a sampling procedure with replacement can be carried out (e.g.\ random vertex sampling with replacement) but then only distinct sampled objects be kept for inference under ``sampling without replacement,'' with the latter referring rather to how collected information is used. In fact, one could develop an analogous inversion approach for such sampling schemes as well, with their own probability matrices $P_{s}$. But we found ``sampling without replacement'' inference for these sampling schemes to be slightly inferior to those with replacement  when all sampled information collected without replacement is used. To be more specific, consider the first sample obtained through vertex sampling with replacement but where only distinct vertices are kept for inference, and the second sample where all vertices are kept for inference. For these two samples, we found inference for the second to be slightly superior. This probably should not be too surprising but might also go against some of the current practices (e.g.\ 
\cite{zhang:etal:2015}) where inference is made on sample graph (see discussion below Eq.\ (\ref{e:P-WR-RES})) 
even if vertices/edges repeat in sampling with replacement as when using a random walk. For the above reason and for  simplicity sake, we decided not to include sampling schemes with replacement where only distinct sampled objects are kept for inference.
\end{remark}

\subsubsection*{\bf RWS}
The inversion results  of this section for RVS and RES can be used in the case of RWS1 and  RWS2/3, respectively, since vertices and edges are sampled uniformly at random in their ``stationary''   regimes.

\subsection{Penalized inversion}
\label{s:inversion-pen}

The performance quality of the estimator (\ref{e:Dj-est-NR}) is rather poor in our sampling settings because the condition numbers of the matrices $P_s$ are generally quite large. See Section 2.1 in 
\cite{zhang:etal:2015} for a related discussion. For example, the condition number of the matrix $P_{s}$ in (\ref{e:P-WR-RNS}) is  approximately $4.1\times 10^{74}$  when $N_v=40,000$, $n_v=2,992$ and $J=185$ (see also Section \ref{s:simulation}).

The quality of the estimation can be improved by considering a penalized estimation. We follow here the constrained, penalized weighted least-squares approach for this problem considered in 
\cite{zhang:etal:2015}, Section 3, used for undirected graphs and different sampling methods.
 More specifically, the penalized estimator is defined as the solution to the optimization problem 
\begin{equation}\label{e:est-penalized}
	\mathop{\rm argmin}_{D} (P_sD - \widehat D_s)^T C^{-1} (P_sD - \widehat D_s) + \lambda \| {\mathcal D} D\|_2^2
\end{equation}
subject to (with $D=(D(0),\ldots,D(J))^T$)
\begin{equation}\label{e:est-penalized-cosntraints}
	D(j)\geq 0,\ j=0,\ldots,J,\quad \sum_{j=0}^J D(j) = N_v.
\end{equation}
In the objective function (\ref{e:est-penalized}), $\lambda>0$ is a penalty parameter (whose choice is discussed below), ${\mathcal D}$ is the second-order differencing operator defined as in Eq.\ (3.2) of 
\cite{zhang:etal:2015}, namely,
\begin{equation}\label{e:est-penalized-2nddiff}
	{\mathcal D} = \left(
		\begin{array}{ccccccccc}
			1 & -2 & 1 & 0 & \ldots & 0 & 0 & 0 & 0 \\
			0 & 1 & -2 & 1 & \ldots & 0 & 0 & 0 & 0 \\ 
			\vdots & \vdots & \vdots & \vdots & \ddots & \vdots & \vdots & \vdots & \vdots \\
			0 & 0 & 0 & 0 & \ldots & 1 & -2 & 1 & 0  \\
			0 & 0 & 0 & 0 & \ldots & 0 & 1 & -2 & 1 
		\end{array}
	\right)
\end{equation}
and used to have smoother solutions, and  $C$ is a suitable weight matrix. Following 
 \cite{zhang:etal:2015}, we take
\begin{equation}\label{e:est-penalized-C}
	C = \mbox{diag}(\widehat D_s) + \frac{\max(\widehat D_s)}{20} I.
\end{equation}
The optimization problem (\ref{e:est-penalized})--(\ref{e:est-penalized-cosntraints}) is implemented by rewriting the objective function as a quadratic function of $D$, and then solving it through the function solve.QP in the R package {\it quadprog} \cite{quadprog}. 

The penalty parameter $\lambda$ is chosen based on the SURE (Stein's unbiased risk estimation) method introduced in 
\cite{eldar:2009}. The use of the SURE method in a context analogous to (\ref{e:est-penalized})--(\ref{e:est-penalized-cosntraints}) is explained in 
\cite{zhang:etal:2015}, Section 3.2, which we refer the reader to for more details. We also use the same tuning parameters as discussed in Section 4.1 of 
\cite{zhang:etal:2015}.

\section{Asymptotic approach}
\label{s:asymptotic}

The estimators introduced in Section \ref{s:inversion} perform poorly for the in-degree distribution tail (as already noted in Section  \ref{s:intro}  and can be seen from the numerical results in Section \ref{s:data}). Estimation in the tail could be addressed through an asymptotic approach, which can then  be combined with the inversion to recover the in-degree distribution over its full range. We shall first explain the basic idea behind the asymptotic approach, and then describe two methods inspired by this approach (Section \ref{s:basic}). Some of the constants in one of the methods will be specific to the sampling scheme and will be derived separately for sampling without replacement (Section \ref{s:asymptotic-s-without-r}) and sampling with replacement (Section \ref{s:asymptotic-s-with-r}).

 \subsection{Basic idea and methods} 
 \label{s:basic}
 To explain the basic principle of the asymptotic approach, recall the two random variables $X$ and $X_s$ introduced in Section \ref{s:sampling}  representing the in-degree and sample in-degree  of a ``general'' vertex, respectively. Moreover, recall  the relations (\ref{e:X-v-rand}) and (\ref{e:E-hat-Ds-ds}) relating these random variables to the in-degree and sample in-degree  distributions. Now note that
\begin{equation}
X_s = \sum_{i=1}^X B_i ,  \label{eq:sumBernoulli}
\end{equation}
where $B_i$ are Bernoulli random variables with parameter $p$ corresponding to the probability that one of the $X$ in-edges is sampled. For example, $p=n_v /N_v$ for RVS without replacement. Depending on the sampling scheme, the variables $B_i$ are independent (when sampling with replacement) or dependent (when sampling without replacement).

The relation (\ref{eq:sumBernoulli}) can be expressed as 
\begin{align}
X_s = p X +\sum_{i=1}^X (B_i -p) &= p X \left (1 + \sqrt{\frac{1-p}{pX}} \frac{1}{\sqrt{Xp (1-p)}} \sum_{i=1}^X (B_i -p) \right ) \nonumber \\
&=: p X  \left (1 + \sqrt{\frac{1-p}{pX}} Z_X \right ). \label{eq:Xs}
\end{align}
For large $X$ and independent $B_i$, the term $Z_X$ is expected to be approximately normal and hence $ \sqrt{(1-p)/(pX)} Z_X$ be negligible. A similar conclusion is also expected when $B_i$ are ``weakly'' dependent. More specifically, as long as 
\begin{equation}\label{e:Bound-X}
\frac{1}{\sqrt{pX}} \leq \epsilon \quad  \mbox{or} \quad X \geq \frac{1}{p \epsilon^2} 
\end{equation}
for some prescribed small level $\epsilon$, one would in fact expect that
\begin{equation} 
X_s \approx p X \label{e:Xs-X-rel-approx}
\end{equation}
and also
\begin{equation}
\Pb (X_s > j) \approx  \Pb (pX > j) \label{e:Xs-X-rel-asym}
\end{equation}
for $j \geq 1/(p \epsilon^2)$. In fact, under mild assumptions on $X$ and $B_i$, the asymptotic relation (\ref{e:Xs-X-rel-asym}) as $j \rightarrow \infty$ follows from Theorem 3.2 in 
 \cite{robert:08}.

\smallskip

The relation (\ref{e:Xs-X-rel-asym}) is at the center of our asymptotic approach. It states that the distribution tail of interest $\Pb(X >j)$ (again, recall (\ref{e:X-v-rand})) can be approximated by $\Pb(X_s > p j)$, which can be estimated in practice (again, recall (\ref{e:E-hat-Ds-ds})). We shall, however, also have a number of refinements of  (\ref{e:Xs-X-rel-asym}) that are  needed for our purposes here. These have to do with:
\begin{itemize}
\item[1.] Working with a p.m.f. rather than a complementary CDF in (\ref{e:Xs-X-rel-asym}).
\item[2.] Dealing with the ``flat'' part of the tail of the p.m.f. of in-degrees.
\item[3.] Accounting for the way the sample in-degrees are collected.
\item[4.] Assuming and modeling of a power-law distribution tail.
\end{itemize}
These points are explained next.

\smallskip

Related to the first point above and similar to the inversion approach, we shall work with the in-degree counts $D(j)=N_v \Pb(X=j)$ and $D_s(j')=N_v \Pb(X_s=j')$ (which is equivalent to working with the in-degree distributions). The relation  (\ref{e:Xs-X-rel-asym}) is expected to yield $D_s(j) \approx p^{-1} D (p^{-1} j)$ or  
\begin{equation}
D (j) \approx  p D_s (p j ).  \label{e:D-Ds-asym}
\end{equation}
An explanation and a word of caution here is that by writing (\ref{e:D-Ds-asym}), we treat $D(j)$ and $D_s(j)$ as ``densities'' so that (\ref{e:D-Ds-asym}) follows by differentiating both sides of  (\ref{e:Xs-X-rel-asym}). This ``density'' view will affect how the results are interpreted. More specifically, the resulting estimators of $D(j)$'s should not be treated as what is expected for fixed $j$'s but rather as an estimated ``density'' for $D(j)$'s from which the in-degrees of a given graph were sampled. Note also that this perspective is slightly different from that for the inversion approach where the estimator of $D(j)$ is obtained for that fixed~$j$.

\smallskip

Related to the second point above, most real networks are such that a range of large in-degrees occurs in the network exactly once. That is,
\begin{equation}\label{e:indegree-power-law-treshold}
D (j) = 1, \quad  \mbox{for a range of $j$'s from $\tau$ to $J$}
\end{equation}
and similalrly 
\begin{equation}\label{e:indegree-power-law-treshold-s}
D_s (j') = 1, \quad  \mbox{for a range of $j'$'s from $\tau_s$ to $J'$.}
\end{equation}
From a modeling perspective, it is interesting to estimate $\tau$ and $J$, and thus also to model the behavior (\ref{e:indegree-power-law-treshold}). An assumption similar to (\ref{e:indegree-power-law-treshold}) was also made in 
\cite{Murai:2017} in the context of the degree distribution of undirected graph.

\smallskip

In connection to the third point above, we note that (\ref{e:D-Ds-asym}) cannot be expected to be used to capture (\ref{e:indegree-power-law-treshold}). This has nothing to do with the way (\ref{e:D-Ds-asym}) was derived or its validity but rather with the way the sample in-degrees are collected. That is, note that after sampling, one would similarly have $\widehat D_s (j')=1$ for a range of large $j'$ and, rescaling these twice by $p$ as in (\ref{e:D-Ds-asym}) would not give (\ref{e:indegree-power-law-treshold}) because $p$ also multiplies $D_s(p j)$ in (\ref{e:D-Ds-asym}). This should be contrasted with the sampling of $X_s$ according to (\ref{eq:sumBernoulli}) for independent copies of $X$ and $B_i$'s, assuming $X$ follows the distribution satisfying (\ref{e:indegree-power-law-treshold}). In this case, 
as we checked numerically but will omit details for shortness sake, the relation (\ref{e:D-Ds-asym}) would also recover  the behavior (\ref{e:indegree-power-law-treshold}). Our asymptotic methods will need to account for these observations, by modeling (\ref{e:indegree-power-law-treshold}) separately.

\smallskip

Finally, related to the fourth point above, some of our methods will be based on the assumption that the in-degree distribution has a power-law tail, namely, 
\begin{equation}\label{e:indegree-power-law}
	D(j)  \approx c \alpha j^{-\alpha-1 },  \quad \mbox{for large $j$ up to  $\tau$},  
\end{equation}
where $\alpha>0$ and $\tau$ appears in (\ref{e:indegree-power-law-treshold}). We use the approximation sign  ``$\approx$'' here and below to indicate that our arguments are not completely rigorous mathematically (though that could also be formalized but at the expense of cumbersome technicalities). It is well known that the in-degree counts   in many real networks  follow a power-law behavior (\ref{e:indegree-power-law}) (see e.g.\  
\cite{Leskovec:2006}, 
\cite{Gjoka:2011} and 
\cite{Murai:2017}). 
\smallskip

We next present two methods inspired by the asymptotic approach discussed above: the ASYM  method and the  LINE method. The LINE method will be constructed by assuming the power-law behavior (\ref{e:indegree-power-law}), whereas this assumption will not be made in the ASYM method.

\subsubsection*{\bf{ASYM method:}}
As discussed around (\ref{e:D-Ds-asym}), the asymptotic approach suggests setting
\begin{equation}\label{e:asym-method-1}
\widehat D (j) =  p \widehat D_s (p j) , \quad \mbox{for large $j$ up to $p^{-1} \widehat  \tau_s$}, 
\end{equation}
where $\widehat \tau_s$ is an estimator for $\tau_s$ in (\ref{e:indegree-power-law-treshold-s}), given by  
\begin{equation}\label{e:tau0-est}
	\widehat  \tau_s = \mathop{\rm argmin}_{j'} \{\widehat D_s(j'):\widehat D_s(j')>0\}
\end{equation}
 (if the argmin results in several $j'$'s, the minimum of these is taken for $\widehat \tau_s$).
In order, to account for the fact that the ``flat'' part of the distribution in (\ref{e:indegree-power-law-treshold}) also contributes to $\widehat D_s (j')$ even for  $j'$ up to $\widehat \tau_s$ after sampling, we shall use 
\begin{equation}\label{e:asym-method-2}
\widehat D (j) =  p \left ( \widehat D_s (p j) - \frac{1}{p} \right ), \quad \mbox{for large $j$ up to $p^{-1} \widehat  \tau_s$}.
\end{equation}
The idea here is that through sampling, the flat ``density'' at value $1$ translates into another flat  ``density''  but at value $1/p$: it is this part which is subtracted from $\widehat D_s(pj)$.
For the ``flat'' part of the distribution, we shall  take
\begin{equation}\label{e:asym-method-3}
\widehat D ( j  )  =  \widehat D_s \left ( \widehat \tau_s + \frac{\widehat J_s - \widehat \tau_s}{\widehat J - p^{-1} \widehat \tau_s } (j- p^{-1} \widehat \tau_s)\right ), 
\quad \mbox{for   $j$ from  $p^{-1} \widehat \tau_s$ to $\widehat J$}, 
\end{equation}
where $\widehat J_s$ is the observed largest sample in-degree and $\widehat J$ is an estimator for $J$.  In view of (\ref{e:Xs-X-rel-approx}), we take
\begin{equation}
\widehat J = \frac{\widehat J_s} {p}. \label{e:J-est}
\end{equation}
That is, using (\ref{e:asym-method-3}) we shall just take the ``flat'' part of $\widehat D_s( j' )$ and stretch it over $p^{-1} \widehat \tau_s$ to $\widehat J$. By doing so, the goal is to capture possibly any variability in the ``flat'' part of $D(j)$ that would be reflected in that for $\widehat D_s( j' )$.

 \subsubsection*{\bf{LINE method:}} At the level of CDF, the relation  (\ref{e:Xs-X-rel-asym}) shows that the power-law is preserved after sampling and also vice versa. To be more rigorous (since  (\ref{e:Xs-X-rel-asym}) does not imply an analogous result for the p.m.f.'s in general), we will show that (\ref{e:indegree-power-law}) implies
\begin{equation}\label{e:indegree-s-power-law}
	D_s(j')  = \E \widehat  D_s(j')  \approx C_s(j') c  \alpha {j'}^{-\alpha-1 },  \quad \mbox{for large $j'$ up to  $\tau_s$}  
\end{equation}
and
\begin{equation}\label{e:Cj-C}
C_s(j') \approx C_s, \quad \mbox{for large $j'$.}
\end{equation}
The quantities $C_s(j')$, $C_s$ will depend on the sampling method used. Though note also that in view of (\ref{e:Xs-X-rel-asym}), one expects 
\begin{equation}\label{e:Cs}
C_s=p^\alpha. 
\end{equation}

Consider now some estimates $\widehat {C_s c}$ and $\widehat \alpha$ modeling the in-degree counts satisfying (\ref{e:indegree-s-power-law})--(\ref{e:Cj-C}), namely,
\begin{equation}
\widehat D_s(j')  \approx \widehat{C_s c}  \widehat \alpha {j'}^{-\widehat \alpha-1 },  \quad \mbox{for large $j'$ up to  $\widehat \tau_s$}, 
\end{equation}
where $\widehat \tau_s$ is defined by (\ref{e:tau0-est}). For example, for $\widehat \alpha$, one could use the maximum likelihood estimator of 
\cite{Clauset:2009} that is part of the  R package {\it igraph} \cite{Csardi:2006}. The relation (\ref{e:indegree-power-law}) then suggests that the  tail of the power-law in-degree distribution could be estimated as 
\begin{equation}\label{e:indegree-power-law-estimators-relation-non-log}
	\widehat D(j)  = \frac{ \widehat {C_sc}}{C_s(j)} \widehat \alpha j^{-\widehat \alpha -1 }, \quad \mbox{for large $j$ up to $\widehat \tau$,}
\end{equation}
where the use of $C_s(j)$  instead of  $C_s$ will make the results slightly more precise in practice for moderate $j$.
The estimator $\widehat \tau$ is defined as follows,
%
%
by relating $\tau$ and $\tau_s$, write 
\begin{equation} \label{eq:tau_est}
 \widehat \tau =  K_1 \widehat \tau_s,
\end{equation}
for a  constant $K_1$. Then, we expect from (\ref{e:indegree-power-law})  that 
\begin{equation}
	1= D (  \tau )  \approx  c \alpha \tau^{-\alpha-1 } \approx  c \alpha (K_1 \widehat \tau_s)^{-\alpha-1 } \approx \frac{K_1^{-\alpha-1 }}{C_s(\widehat \tau_s)}  \widehat D_s (\widehat \tau_s) 
\end{equation}
and, hence,
\begin{equation}\label{eq:K1}
	 K_1   \approx      \left ( \frac{1 }{C_s(\widehat \tau_s)}  \widehat D_s(\widehat \tau_s) \right )^{1/(\widehat \alpha+1) }.
\end{equation}

We also need to define an estimator for the behavior (\ref{e:indegree-power-law-treshold}), which we take as
\begin{equation}\label{e:est-Dj-eq-1}
\widehat D (j) =1, \quad \mbox{for $j$ from $\widehat \tau$ to $\widehat J$}. 
\end{equation}
where $\widehat J$ is defined as in (\ref{e:J-est}).
Following (\ref{e:D-Ds-asym}), by writing (\ref{e:est-Dj-eq-1}), we actually view $\widehat D (j)$ as a  ``density''  for the specified values of $j$, that is, we do not claim that $D(j)=1$ for all $j$ in the range (or how many $j$'s are such). If needed, this ``density'' could be normalized to 1, when it is produced over the whole range of $j$ from 0 to $\widehat J$.\\

\begin{remark}
In view of (\ref{e:Bound-X}), the ASYM method is expected to be particularly powerful when the largest in-degree $J$ is large. Indeed, for example, with $\epsilon=0.1$ and $p=0.1$, note that $(\ref{e:Bound-X})$ becomes $X\geq 1000$. Furthermore, we stress again that the ASYM method applies to any distribution tail, whereas the LINE method to a power-law tail only. On the other hand, the LINE method seems to work well even for smaller $J$, whenever a power-law tail is present.
\end{remark}

The following sections will provide the forms of the quantities $C_s(j')$, $C_s$   in (\ref{e:indegree-s-power-law})--(\ref{e:Cj-C}).

\subsection{LINE method: sampling without replacement}
\label{s:asymptotic-s-without-r}
As in Section \ref{s:inversion-woR}, we consider separately the RVS and RES schemes.


%

\subsubsection*{\bf{RVS}}

We will show that  
\begin{multline}
C_s (j') =  \left (\frac{n_v}{N_v} \right )^\alpha \Big ( 1+ \frac{a -2 \alpha-1}{2 j' a}\Big )^{a(j' +1/2)-\alpha -1} \\
	\times  \Big ( 1+ \frac{a -\alpha}{a n_v} \Big )^{\alpha-a(1+n_v) +1/2}  \Big ( 1 + \frac{a+1}{2a(n_v-j') }  \Big )^{a (n_v - j'+1/2)} , \label{eq:Cj-RNS-NR}
\end{multline}
where $a=1/(1-n_v/N_v)$, which for large $j'$ and $n_v$ approaches $C_s =p^\alpha=  \left(\frac{n_v}{N_v} \right )^\alpha$ as in (\ref{e:Cs}). 
Indeed, 
observe from  (\ref{e:E-hat-Ds-matrix}) and (\ref{e:P-NR-RNS}) that 
\begin{equation}\label{e:asy-est-EDNR0}
D_{s}(j') =	\E \widehat D_{s}(j')   \approx  c \alpha    \sum_{j=j'}^{N_v -1}  \frac{\binom{j}{j'} \binom{N_v-j}{n_v-j'}} { \binom{N_v}{n_v}} j^{-\alpha-1},
	\quad  \mbox{for large $j'$}.
\end{equation}
By using the normal approximation to the hypergeometric distribution (
\cite{Feller:1968}, p.\ 194, Eq.\ (7.5)), we have
\begin{align}
 \frac{\binom{j}{j'} \binom{N_v-j}{n_v-j'}} { \binom{N_v}{n_v}} & \approx \frac{1} {\sqrt {  \frac{2 \pi n_v j}{N_v} \left (1- \frac{j}{N_v} \right ) (1-t) }}
   \exp \left \{ -a \left ( \left (j' -  \frac{n_v j}{N_v} \right)  \bigg / \sqrt{   \frac{2 n_v  j}{N_v} \left (1-\frac{j}{N_v} \right )}\right )^2 \right \}  \nonumber \\  
 & =\varphi \left (j'; \frac{n_v j}{N_v},\frac{n_v  j}{N_v} \left (1-\frac{j}{N_v} \right ) \right )^a \frac{\sqrt{  \left (\frac{2 \pi n_v j}{N_v} \left (1-\frac{j}{N_v} \right ) \right )^{a-1} }} {\sqrt{1-t}}, \label{eq:normal_aprox}
\end{align}
where  $a=1/(1-n_v/N_v)$, $t=n_v/N_v$ and $\varphi(x; \mu,\sigma^2)$ is the  p.d.f.  of the normal distribution with mean  $\mu$ and variance $\sigma^2$.
 In order to proceed with the asymptotic analysis, we shall replace $\varphi(j'; n_v j/N_v,n_v  j/N_v (1-j/N_v)$ with the p.m.f.~of the binomial distribution 
 with parameters $n_v$ and $j/N_v$ (see also  Remark \ref{r:binomial-approx} below).
 Therefore, by substituting  (\ref{eq:normal_aprox}) in (\ref{e:asy-est-EDNR0}) and using the binomial replacement, we have

\begin{align}
	\E &  \widehat D_{s}(j')   \sim c \alpha N_v^{-\alpha-1} \sqrt{\frac{(2 \pi n_v)^{a-1} }{1-t}}  {\binom{n_v}{j'}}^a 
	 \sum_{j=j'}^{N_v -1}   \left (\frac{j}{N_v} \right )^{j'a +\frac{a-1}{2}- \alpha -1} \left (1- \frac{j}{N_v} \right )^{(n_v -j')a +\frac{a-1}{2}}
	 \nonumber \\ 
	&  \approx c \alpha N_v^{-\alpha} \sqrt{\frac{(2 \pi n_v)^{a-1} }{1-t}}  {\binom{n_v}{j'}}^a 	
	  \int_{j'/N_v}^{1}   x^{j'a +\frac{a-1}{2}- \alpha -1}  (1-x )^{(n_v -j')a +\frac{a-1}{2}} \, dx 
	   \nonumber \\ 
	 &  \approx c \alpha N_v^{-\alpha} \sqrt{\frac{(2 \pi n_v)^{a-1} }{1-t}}  {\binom{n_v}{j'}}^a 
	\Big [ B \Big (1,j'a +\frac{a-1}{2}- \alpha ,(n_v -j')a +\frac{a-1}{2}+1 \Big) 
	 \nonumber \\ 
	 & \hspace{6cm}-B \Big (j'/N_v, j'a +\frac{a-1}{2}- \alpha ,(n_v -j')a +\frac{a-1}{2}+1 \Big ) \Big ] 
	  \nonumber \\ 
	 &	 \approx c \alpha N_v^{-\alpha} \sqrt{\frac{(2 \pi n_v)^{a-1} }{1-t}}  {\binom{n_v}{j'}}^a  
	B \Big (1,j'a +\frac{a-1}{2}- \alpha ,(n_v -j')a +\frac{a-1}{2}+1 \Big ) 
	 \nonumber \\ 
	 & \approx  c \alpha   { j'}^{-\alpha-1} \left (\frac{n_v}{N_v} \right )^\alpha \Big ( 1+ \frac{a -2 \alpha-1}{2 j' a}\Big )^{a(j' +1/2)-\alpha -1} \nonumber
	 \Big ( 1+ \frac{a -\alpha}{a n_v} \Big )^{\alpha-a(1+n_v) +1/2} \\
	 & \hspace{8.5cm} \times \Big ( 1 + \frac{a+1}{2a(n_v-j') }  \Big )^{a (n_v - j'+1/2)}  
	 \label{eq:e:asy-est-EDNR0-aux}  \\
	&\approx  c \alpha   { j'}^{-\alpha-1} \left (\frac{n_v}{N_v} \right )^\alpha, \label{eq:e:asy-est-EDNR0-final}
\end{align}
where (\ref{eq:e:asy-est-EDNR0-aux}) follows from 
$\binom{n}{k}  \sim \frac{n^{n+1/2} }{ \sqrt{2 \pi}  k^{k+1/2} (n -k)^{n-k+1/2}}$ and
$B(x,y) \sim \sqrt{2 \pi} \frac{x^{x-1/2} y^{y-1/2} }{(x+y)^{x+y-1/2}}$ by using  Stirling's approximation. 
The relations  (\ref{eq:e:asy-est-EDNR0-final}) and  (\ref{eq:e:asy-est-EDNR0-aux}) justifies the form of $C_s(j')$ in (\ref{eq:Cj-RNS-NR})
 and $C_s$, respectively.


\begin{remark}
\label{r:binomial-approx}
 In the argument given above, we approximated a hypergeometric distribution by a normal distribution which we then
replaced by a binomial distribution. In fact, by using directly another well-known  binomial approximation to the hypergeometric distribution (see e.g.\ 
\cite{Feller:1968}, p.\ 172, Problem 35), 
we obtain different $C_s(j')$ that perform worse in practice and the same constant $C_s$.
\end{remark}

\subsubsection*{\bf{RES}}

Similarly, we have that
\begin{multline}
C_s(j') =  \left (\frac{n_e}{N_e} \right )^\alpha \Big ( 1+ \frac{a -2 \alpha-1}{2 j' a}\Big )^{a(j' +1/2)-\alpha -1} \\
	 \times \Big ( 1+ \frac{a -\alpha}{a n_e} \Big )^{\alpha-a(1+n_e) +1/2}  \Big ( 1 + \frac{a+1}{2a(n_e-j') }  \Big )^{a (n_e - j'+1/2)}, \label{eq:Cj-RES-NR}
\end{multline}
where $a=1/(1-n_e/N_v)$, which for large $j'$ and $n_e$ approaches  $C_s = p^\alpha= \left(\frac{n_e}{N_e} \right )^\alpha$.  
Indeed, 
 observe from  (\ref{e:E-hat-Ds-matrix}), (\ref{e:P-NR-RES}) and (\ref{e:indegree-power-law}) that 
\begin{equation}
D_{s} (j')=	\E \widehat D_{s} (j')   \approx  c \alpha    \sum_{j=j'}^{N_e}  
	\frac{\binom{j}{j'} \binom{N_e-j}{n_e-j'}} { \binom{N_e}{n_e}}j^{-\alpha-1}
\end{equation}
which implies   (\ref{eq:Cj-RES-NR}) 
 as in (\ref{e:asy-est-EDNR0})--(\ref{eq:e:asy-est-EDNR0-aux}). 


\subsection{LINE method: sampling with replacement}
\label{s:asymptotic-s-with-r}
As in Section \ref{s:inversion-wR}, we consider separately the RVS, RES and RWS schemes. 



\subsubsection*{\bf{RVS}}

We will show that $C_s(j')$ 
is given by
\begin{equation}\label{e:RVS-Cj}
C_s(j') = \left (\frac{n_v}{N_v} \right )^\alpha  e \Big ( 1 - \frac{\alpha+1}{j'} \Big )^{j'-\alpha-1/2}   \Big ( 1 - \frac{\alpha}{n_v} \Big )^{\alpha-n_v +1/2} 
\end{equation}
and for large $j'$ and $n_v$ approaches 
	$C_s = p^\alpha=   \left ( \frac{n_v}{N_v}\right )^\alpha$. 
Indeed,
\begin{align}
D_{s} (j') =	\E \widehat D_{s}(j')  & \approx  c \alpha  \binom{n_v}{ j'} \sum_{j=1}^{N_v -1} \left (\frac{j}{N_v} \right )^{ j'}  \left (1- \frac{j}{N_v} \right )^{n_v - j'} j^{-\alpha-1} \nonumber \\
	& \approx  c \alpha N_v^{-\alpha}  \binom{n_v}{j'}  \int_{0}^{1} x^{ j'-\alpha-1} (1-x)^{n_v- j'} \, dx \nonumber \\
	& =  c \alpha N_v^{-\alpha}  \binom{n_v}{ j'}   B ( j'-\alpha,n_v-  j'+1) \nonumber \\
&\approx  c \alpha {j'}^{-\alpha-1}  \left (\frac{n_v}{N_v} \right )^\alpha  e \Big ( 1 - \frac{\alpha+1}{j'} \Big )^{j'-\alpha-1/2}   \Big ( 1 - \frac{\alpha}{n_v} \Big )^{\alpha-n_v +1/2}   \\
 &\approx  c \alpha {j'}^{-\alpha-1}  \left (\frac{n_v}{N_v} \right )^\alpha \label{e:asy-est-DWR0} .
\end{align}

\subsubsection*{\bf{RES}}
We have that 
\begin{equation}\label{eq:Cj-RES-WR}
C_s(j') = \left (\frac{n_e}{N_e} \right )^\alpha  e \Big ( 1 - \frac{\alpha+1}{j'} \Big )^{j'-\alpha-1/2}   \Big ( 1 - \frac{\alpha}{n_e} \Big )^{\alpha-n_e +1/2}  
\end{equation}
 which for large $j'$ and $n_e$  is approximately $C_s =  p^\alpha=  \left ( \frac{n_e}{N_e}\right )^\alpha$. 
Indeed, 
observe from  (\ref{e:E-hat-Ds-matrix}), (\ref{e:P-WR-RES}) and (\ref{e:indegree-power-law}) that
\begin{align}
D_{s} (j') =	\E \widehat D_{s}(j')  & \approx  c \alpha  \binom{n_e}{ j'} \sum_{j=1}^{N_e} \left (\frac{j}{N_e} \right )^{ j'}  \left (1- \frac{j}{N_e} ,\right )^{n_e - j'} j^{-\alpha-1} , 
\end{align}
which then can be dealt with by  the same reasoning as in the derivation of (\ref{e:asy-est-DWR0}).


\subsubsection*{\bf{RWS}} The asymptotic results  of this section for RVS and RES can be used in the case of RWS1 and RWS2/3, respectively.

\section{Data study}
\label{s:data}

\begin{figure}[t!]
  \begin{center}
      \subfigure[RVS-WR]{\label{f:RVS-WR}\includegraphics[scale=0.51]{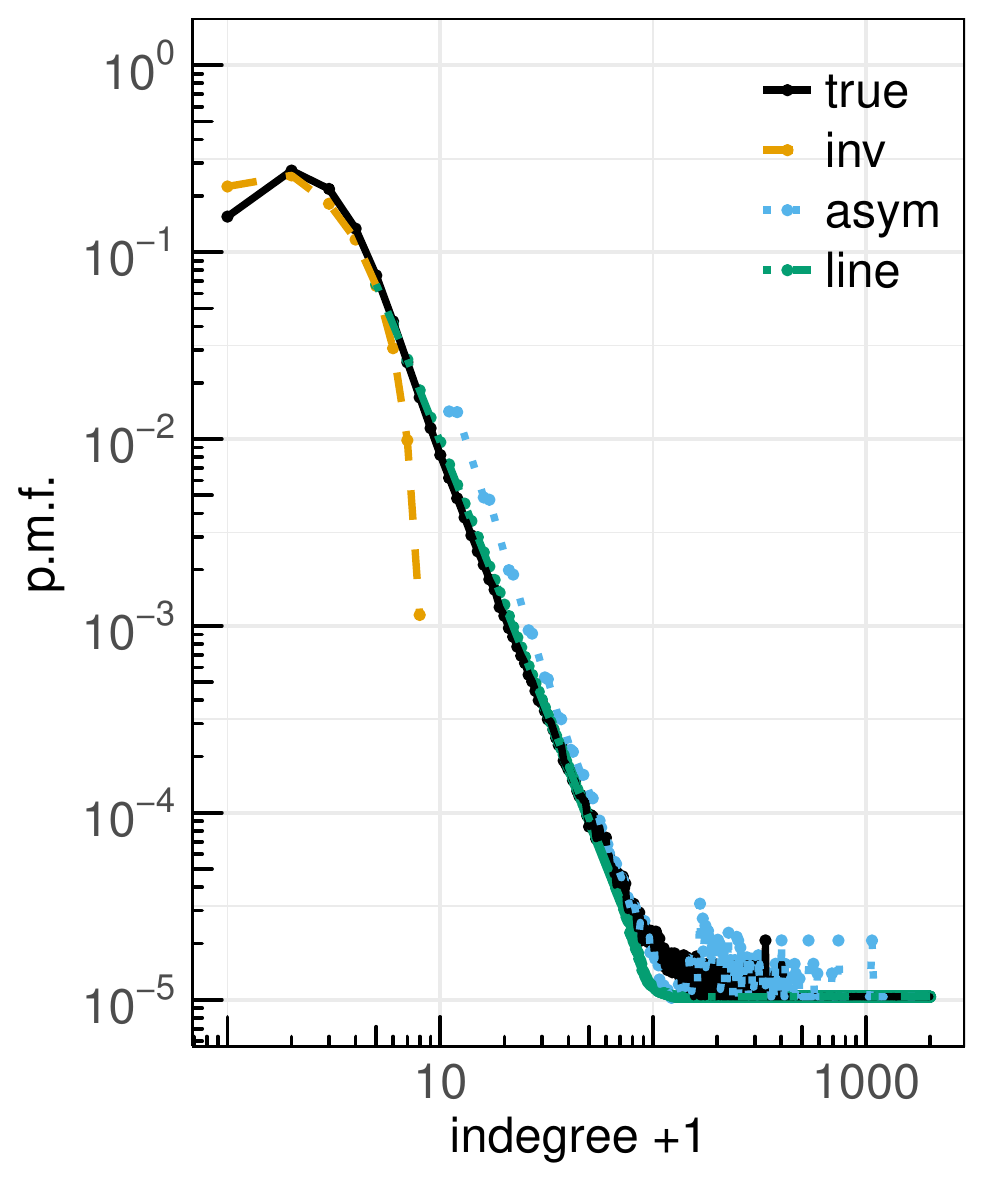}}
      \subfigure[RWS1 (jump rate: 30\%)  ]{\label{f:RWS1}\includegraphics[scale=0.51]{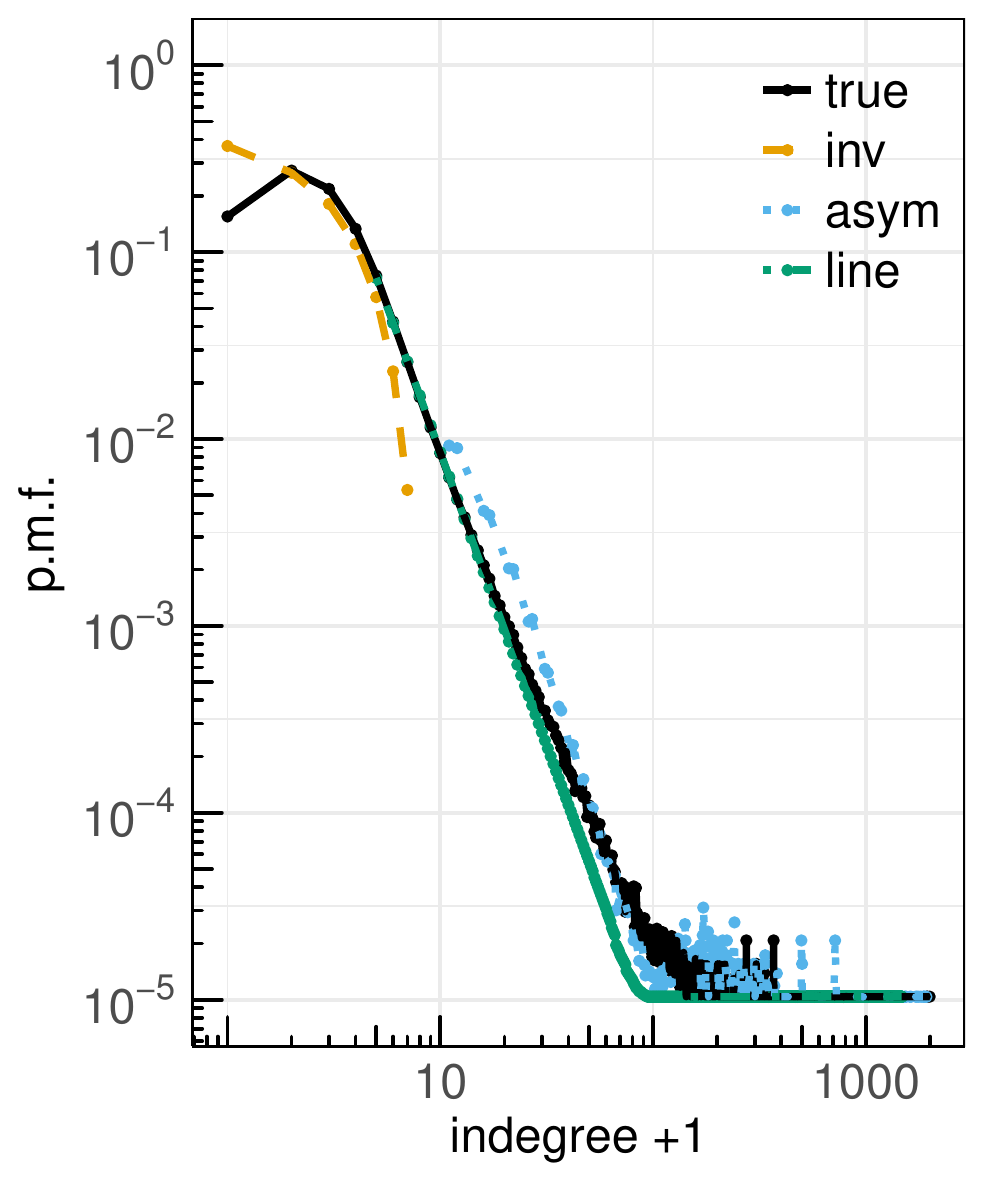}}
  \end{center}
  \caption{Power-law networks  and random vertex based sampling  ($p=0.2$).}
  \label{fig:RVS}
  \vspace{-.5cm}
\end{figure}

In this section, we first present a simulation study  to illustrate and to compare the performances of the estimators introduced  in Sections \ref{s:inversion} and \ref{s:asymptotic}, 
on networks simulated from  random directed graph models. We then examine the proposed estimation methods on two real-world directed networks. The study of this section was performed with the help of  the R package {\it igraph} \cite{Csardi:2006}.

 
\subsection{Normalizations} Under  RVS-WR,  $n_v$ vertices are sampled, with  each one having on  average  $N_e / N_v$  out-edges.
%
%
%
 In order to have approximately the same number of sampled edges as for RES-WR,  we set $n_e=n_v N_e / N_v$. We assume that  $N_v$ and $N_e$ are known or can be estimated  from suitable sample quantities of the sample graph using generalizations of the Horvitz-Thompson estimator
as  e.g.\ in  \cite{Kolaczyk:2009}, Chs.~5.4 and 5.5.
The comparison of RVS-WR and RES-WR against RVS-NR and RES-NR could be performed using a normalization based on a budget with different costs of sampling and  re-sampling a vertex (edge) but this will not be pursued here. 
 Any differences between sampling with and without replacement should be  observable   when the number of sampled vertices (edges) is  large compared to the total number of vertices (edges) 
 (say, $p>0.15$).

 For the proposed RW algorithms  that sample either vertices or edges at random with replacement  (in the limit), we use the same 
 sample size normalization as in RVS-WR and RES-WR, respectively. We also equate the jump probabilities in the different random walks.
 For RWS2, the jump probability, $w/(1+w)$, is constant and does not depend on the degree of the vertices.  For RWS1 and RWS3,
the jump rate can be approximated   at each step $i$ of the constructed undirected graph $G^i$ through $w/(\bar d_i + w)$, where $\bar d_i$  
 is the mean degree of the graph. 
  Finally, we note that  typical sampling proportions $p$ used in the literature vary between $0.1$ and $0.3$.




\subsection{Synthetic directed networks}
\label{s:simulation}

%
%
%
We generated 30 directed  scale-free networks 
where the in-degree and the out-degree distributions  follow a power law. The expected numbers of nodes, edges,  in-degree  exponent and out-degree  exponent,  are $10^{5}$, $3\times 10^5$, $1.5$ and $1.5$, respectively.  For each directed network, we estimated the in-degree distribution in the largest component using the inversion and asymptotic approaches under   sampling methods  with replacement. The proportion $p$ of vertices (edges) sampled is $0.2$ and the jump rate for the RWS algorithms is set to 30\%.

Figures \ref{f:RVS-WR}--\ref{f:RWS1} show the estimation results for  RVS-WR and RWS1, respectively, where vertices are sampled uniformly (in the limit for the RW).
The  ``true'' line corresponds to the average of the p.m.f.'s  of the 30 generated networks. The average of the estimates from the inversion approach with penalization (\ref{e:est-penalized})--(\ref{e:est-penalized-cosntraints}), labeled  ``inv,'' can recover the beginning (bulk)  of the distribution with some bias for the in-degree zero.  We note that without penalization, the variance of the unbiased estimator (\ref{e:Dj-est-NR}) is so large  that   the estimator of  the distribution is impractical. The penalization parameter $\lambda$ in (\ref{e:est-penalized}) has  the effect of shrinking the estimates of the in-degree distribution,  that leads to a substantial reduction in the variance, at the expense of  increasing the  bias.  This approach also controls the variance in the tail by  forcing  the estimates to be equal to zero.  

For the asymptotic approach, we plot in Figures \ref{f:RVS-WR}--\ref{f:RWS1}  the (average) estimates given by the ASYM method (\ref{e:asym-method-2})--(\ref{e:asym-method-3}) and LINE method (\ref{e:indegree-power-law-estimators-relation-non-log}) (along with (\ref{e:RVS-Cj})) and (\ref{e:est-Dj-eq-1}) which allow recovering the tail of the distribution  complementing the inversion approach. 
Since the LINE method is sensitive to the estimate of $\alpha$ in  (\ref{e:indegree-power-law-estimators-relation-non-log}), we estimate it  from the ``true'' distribution, in order to be able to detect differences between the sampling methods (in Section \ref{s:RealWorldDirectGraphs} below, we use the sample in-degree distribution which satisfies (\ref{e:indegree-s-power-law}) and is the only one available in practice).
The ASYM method is  accurate for larger in-degrees while the  LINE method works well even for smaller in-degrees; however, the latter can only be applied to a power-law tail. The comparison  between Figures \ref{f:RVS-WR} and \ref{f:RWS1} shows that RWS1  with a moderate jump rate can approximate RVS-WR (which can be viewed as RWS1 with jumps only).

\begin{figure}[t]
  \begin{center}
      \subfigure[RES-WR]{\label{f:RES-WR}\includegraphics[scale=0.51]{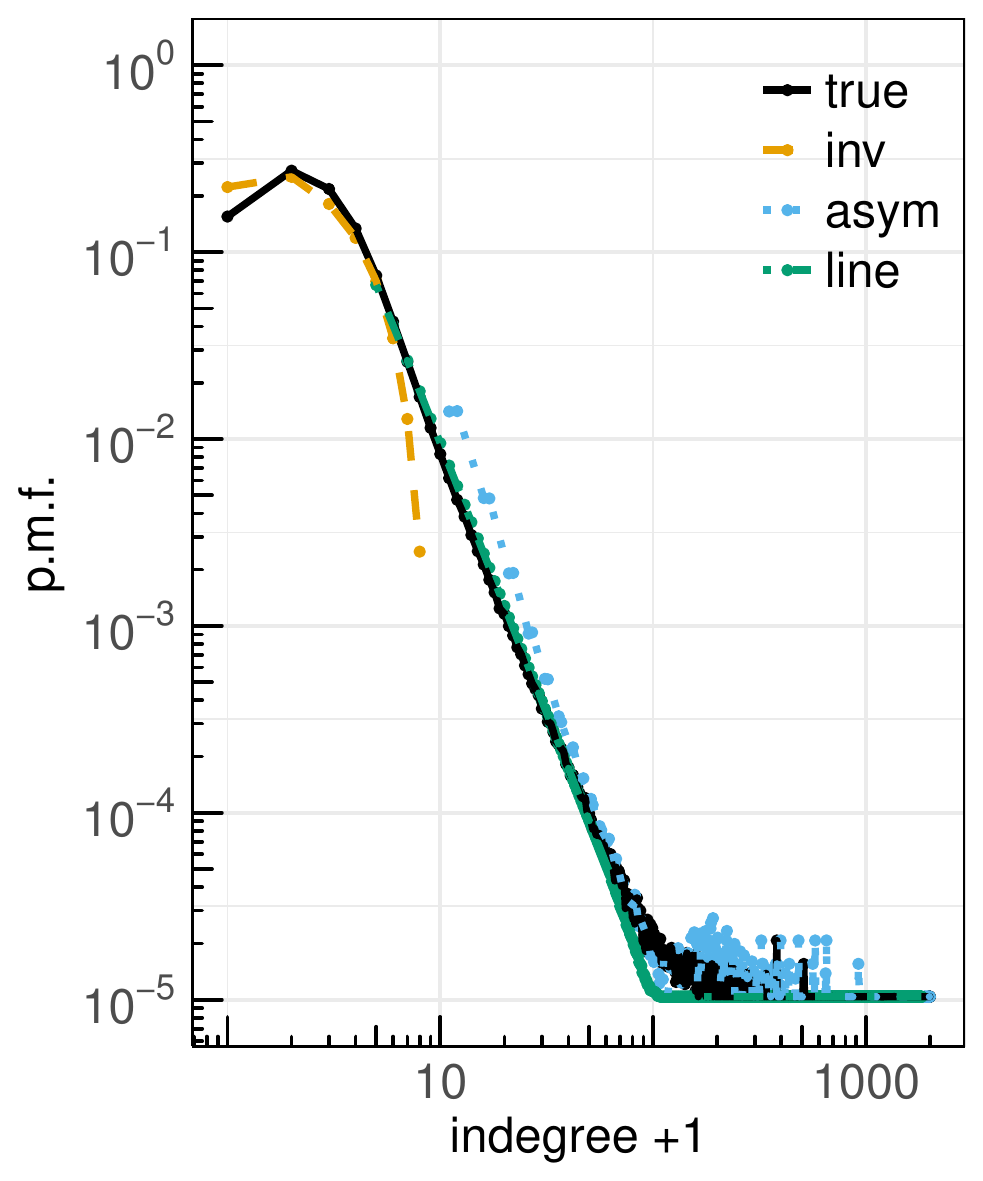}}
      \subfigure[RWS2 (jump rate: 30\%)]{\label{f:RWS2}\includegraphics[scale=0.51]{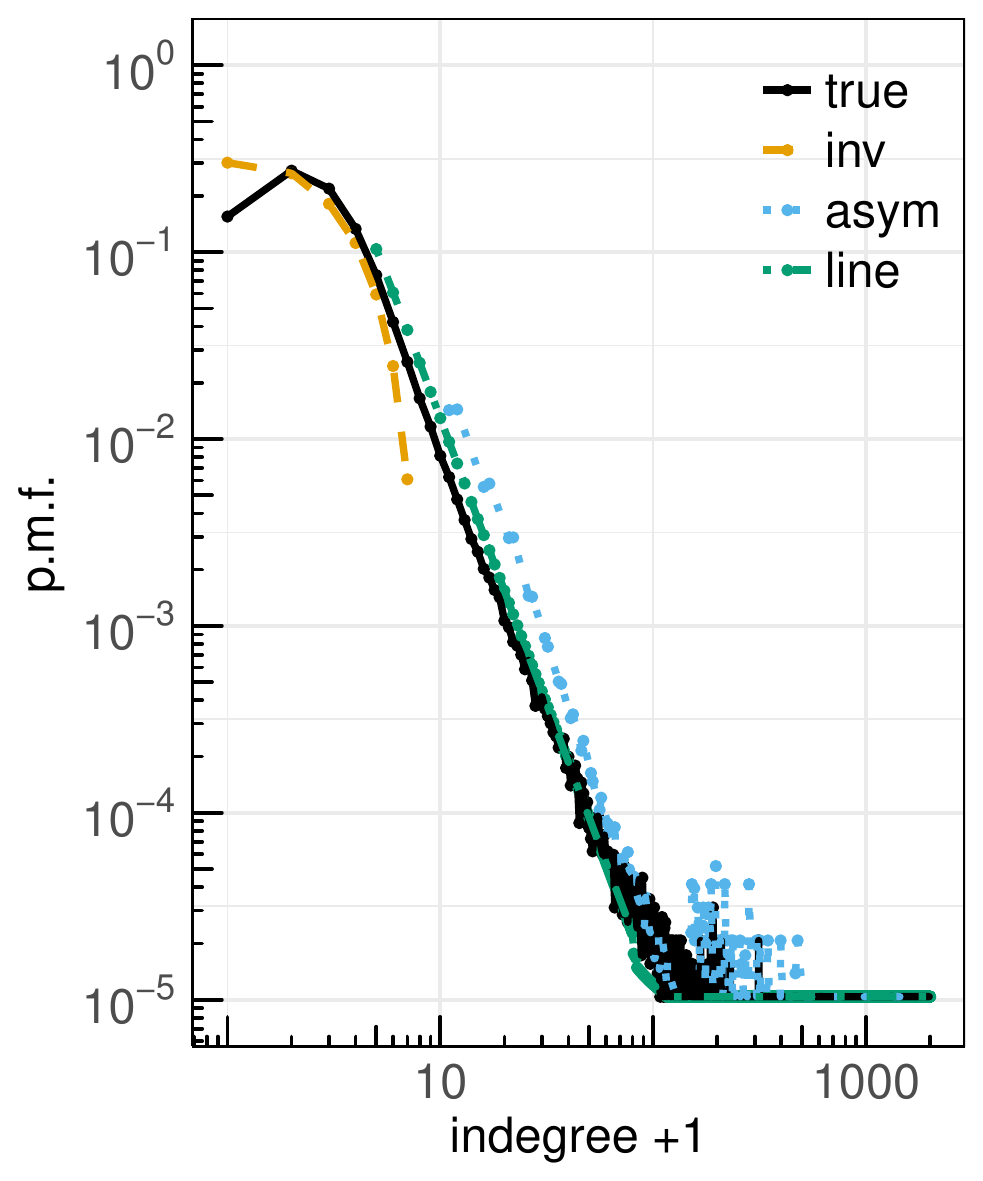}}
      \subfigure[RWS3 (jump rate: 30\%)]{\label{f:RWS3}\includegraphics[scale=0.51]{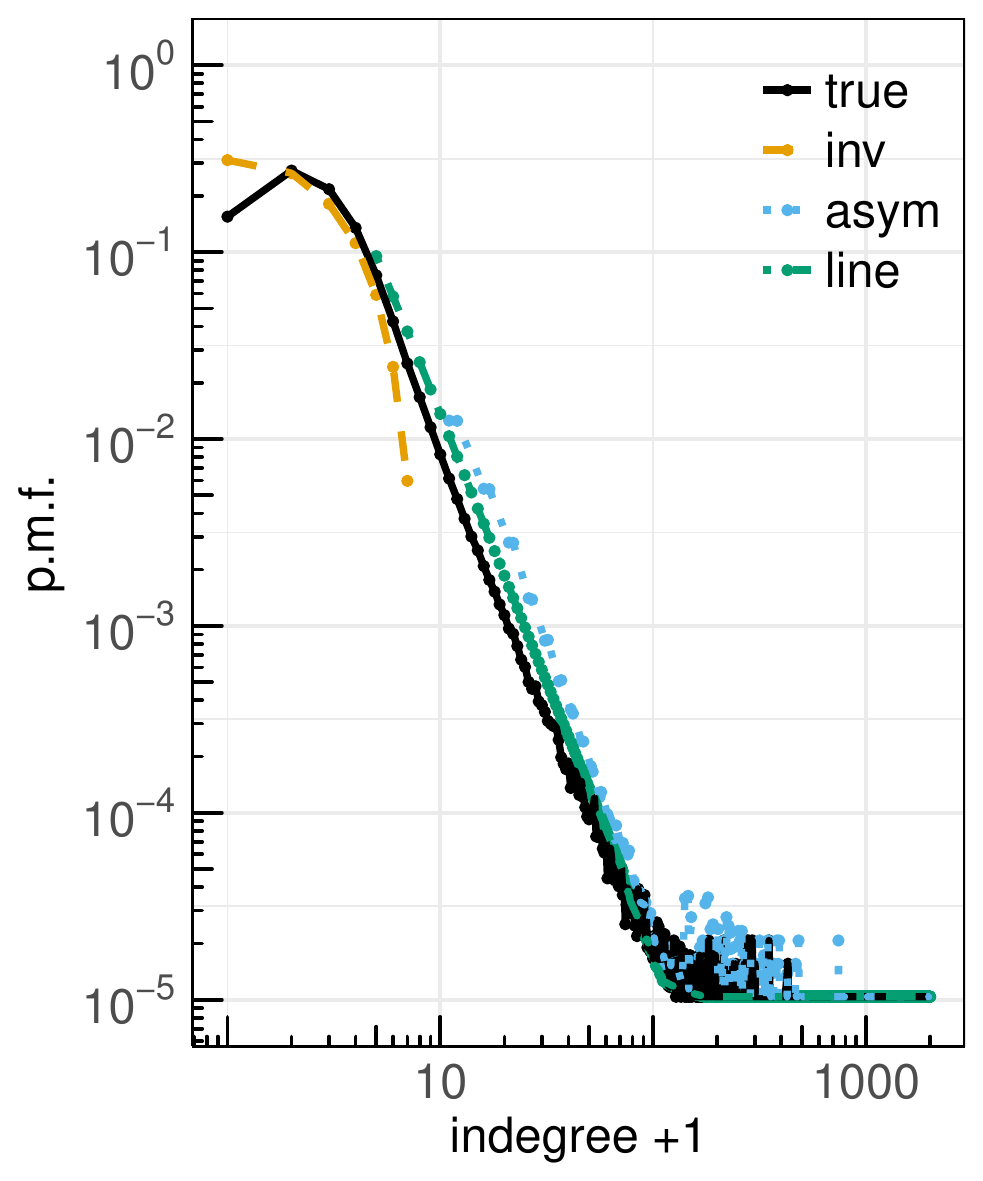}}   
  \end{center}
  \caption{Power-law networks and random edge based sampling ($p= 0.2$).}
  \label{fig:edge}
\end{figure}

\begin{figure}[t]
  \begin{center}
      \subfigure[RVS-WR]{\label{f:RNS-WR}\includegraphics[scale=0.51]{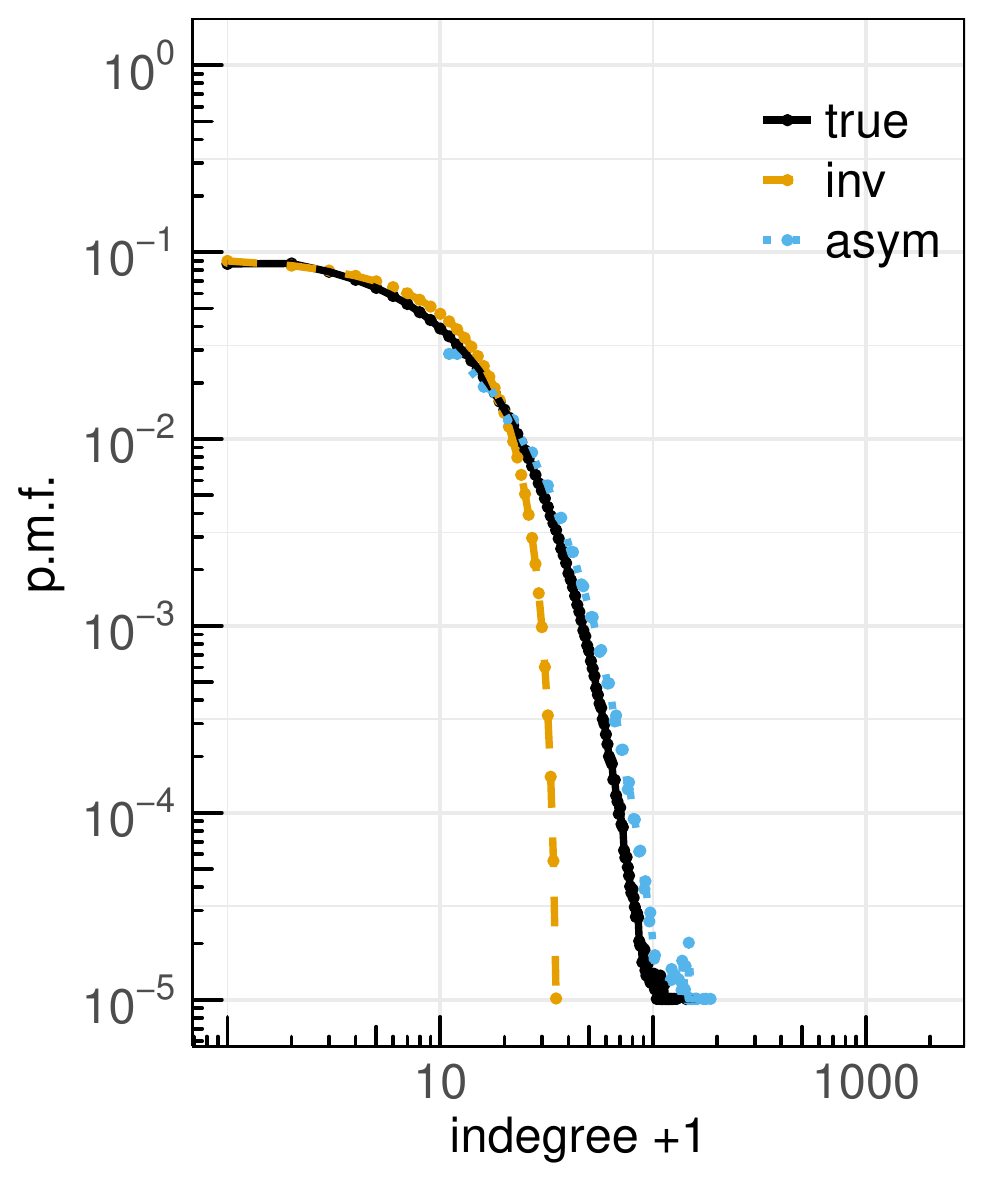}}
      \subfigure[RWS1 (jump rate: 15\%)]{\label{f:RNS-WR}\includegraphics[scale=0.51]{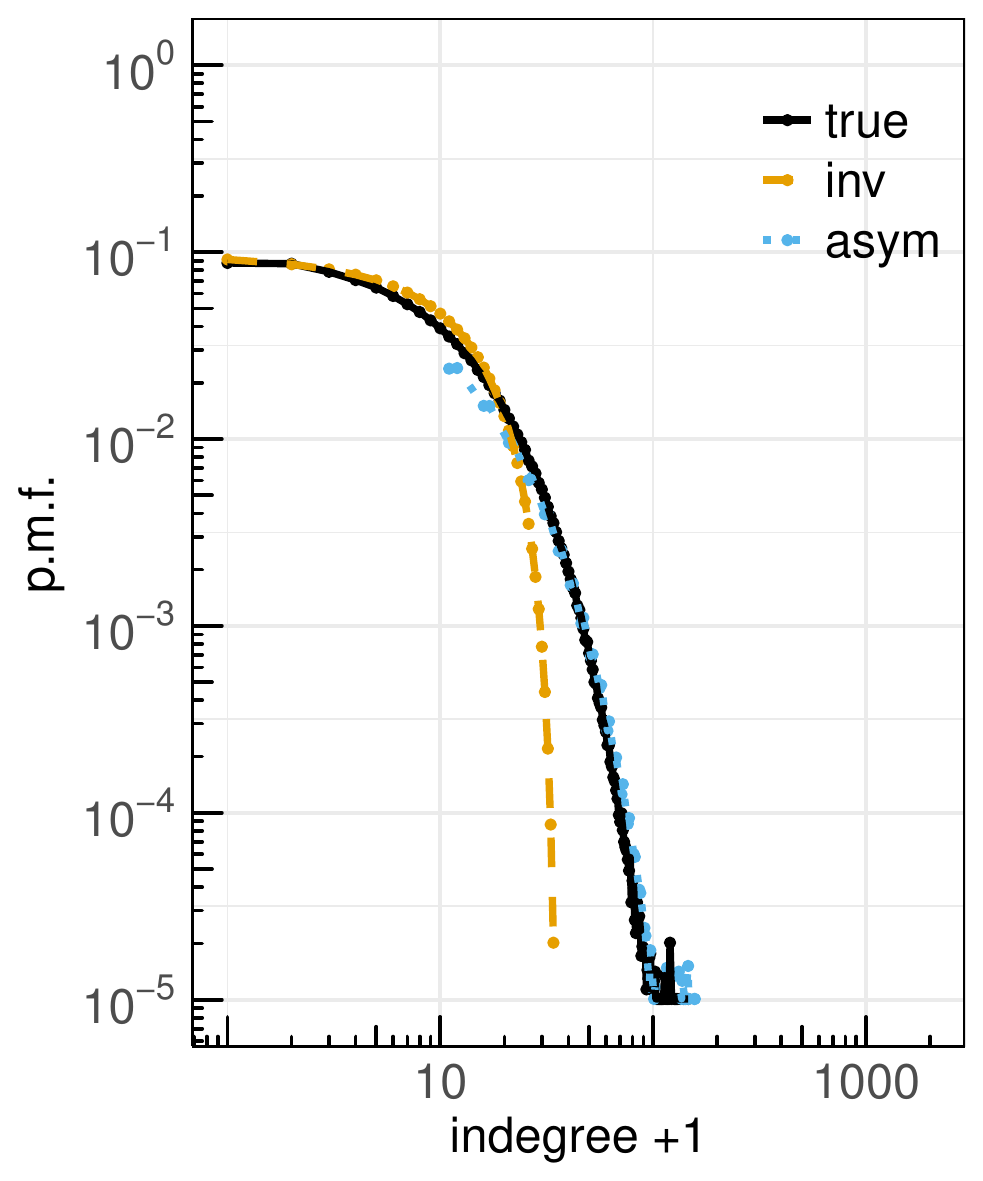}}
  \end{center}
  \caption{Non power-law network and random vertex based sampling $(p=0.15)$.}
  \label{fig:non-power-law}
\end{figure}

\begin{figure}[t]
  \begin{center}
          \subfigure[\tiny RWS1(jump rate: 30\%)]{\label{f:RWS1-real}\includegraphics[scale=0.38]{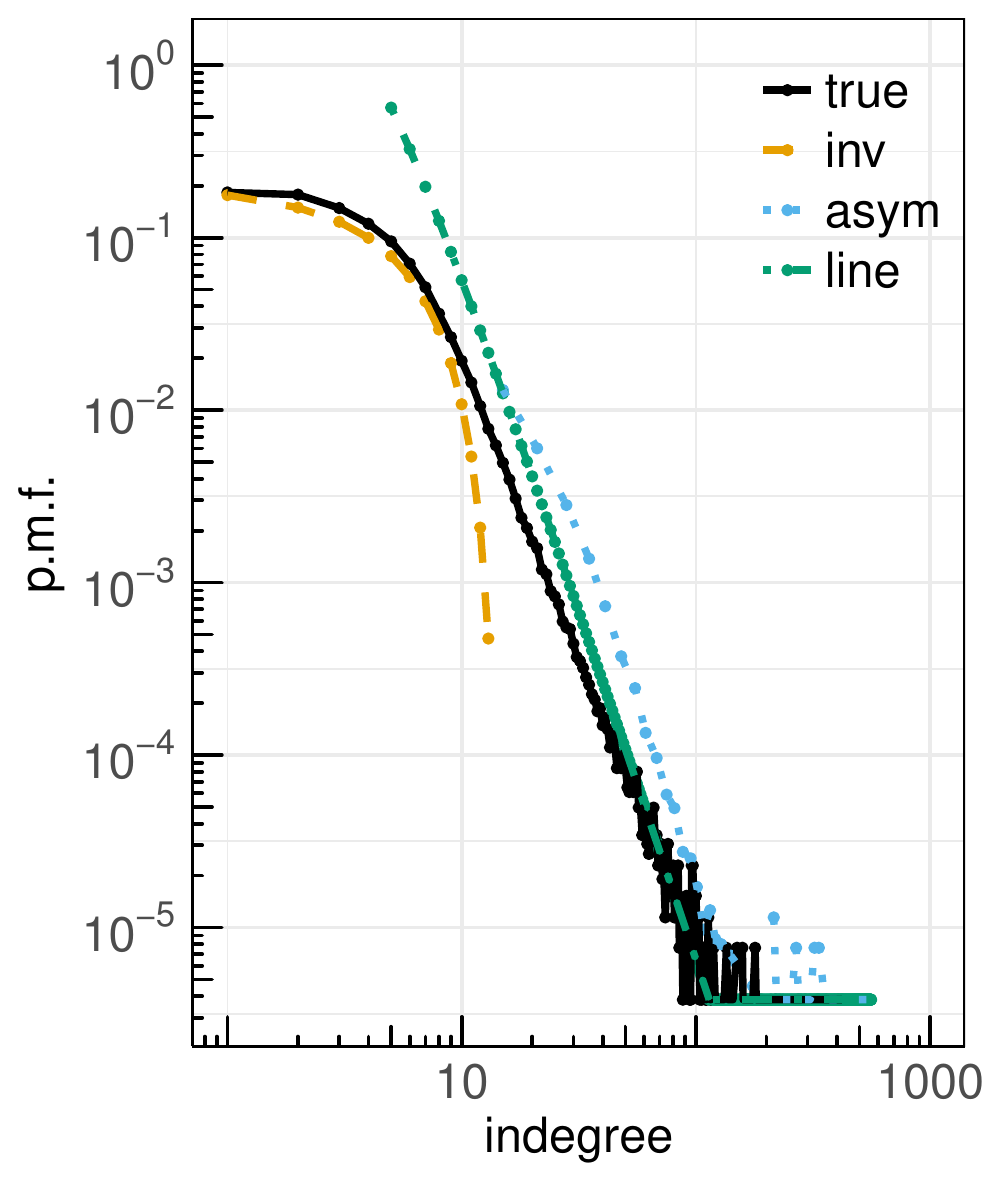}}
        \subfigure[\tiny RWS2 (jump rate: 30\%)]{\label{f:RWS2-real}\includegraphics[scale=0.38]{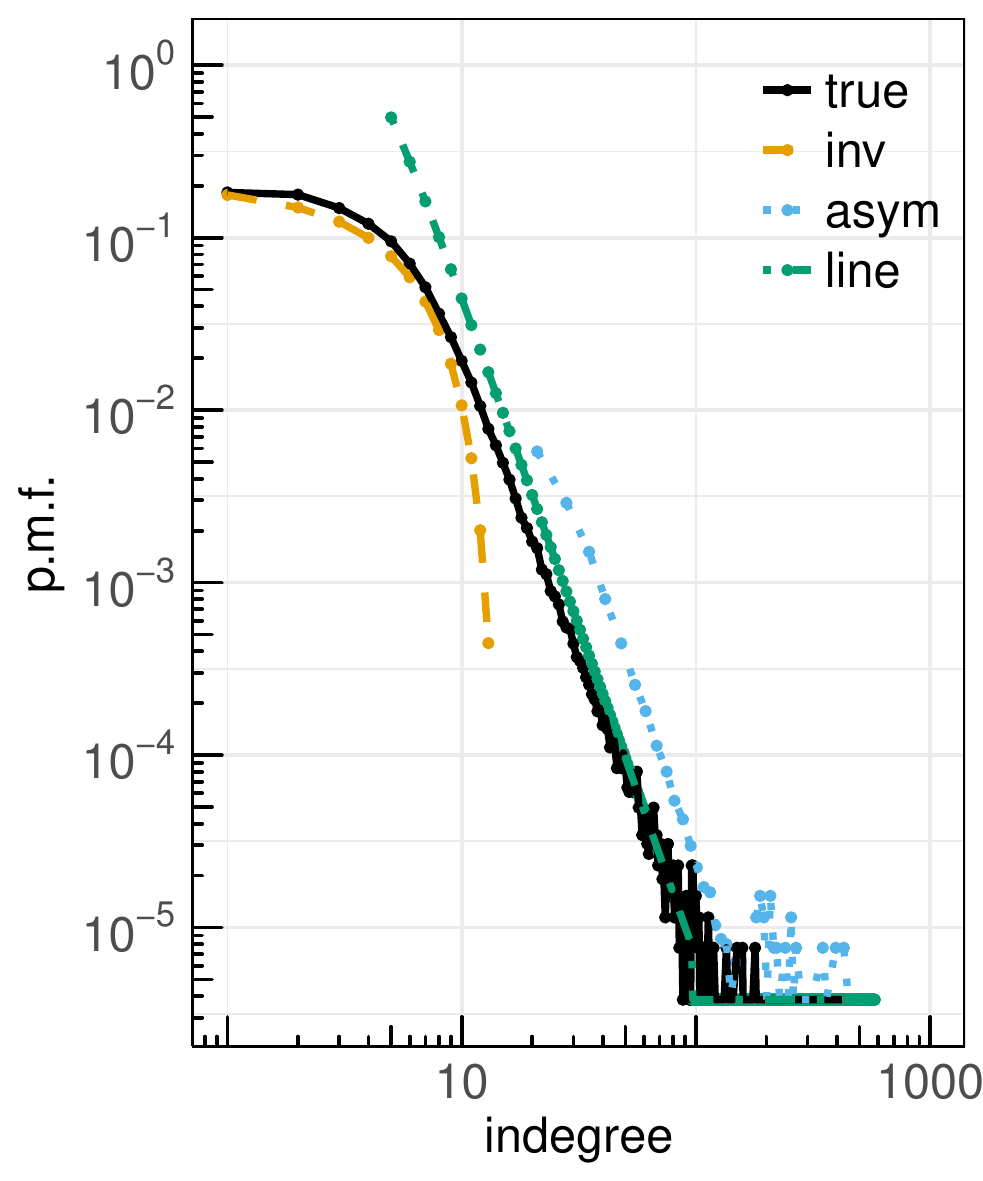}}  
        \subfigure[\tiny RWS3 (jump rate: 30\%)]{\label{f:RWS3-real}\includegraphics[scale=0.38]{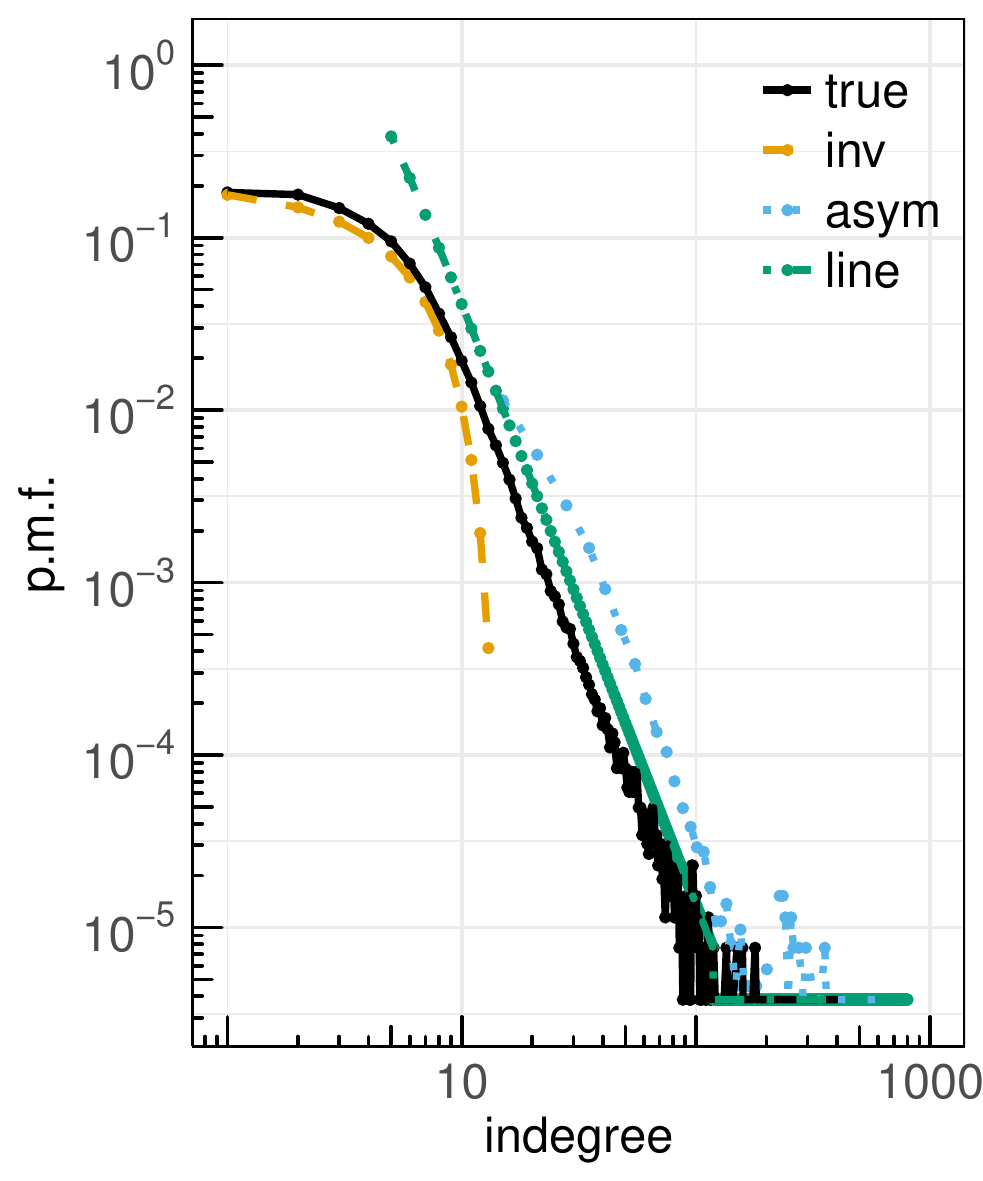}}  
        \subfigure[\tiny RVS-WR]{\label{f:RVS-real}\includegraphics[scale=0.38]{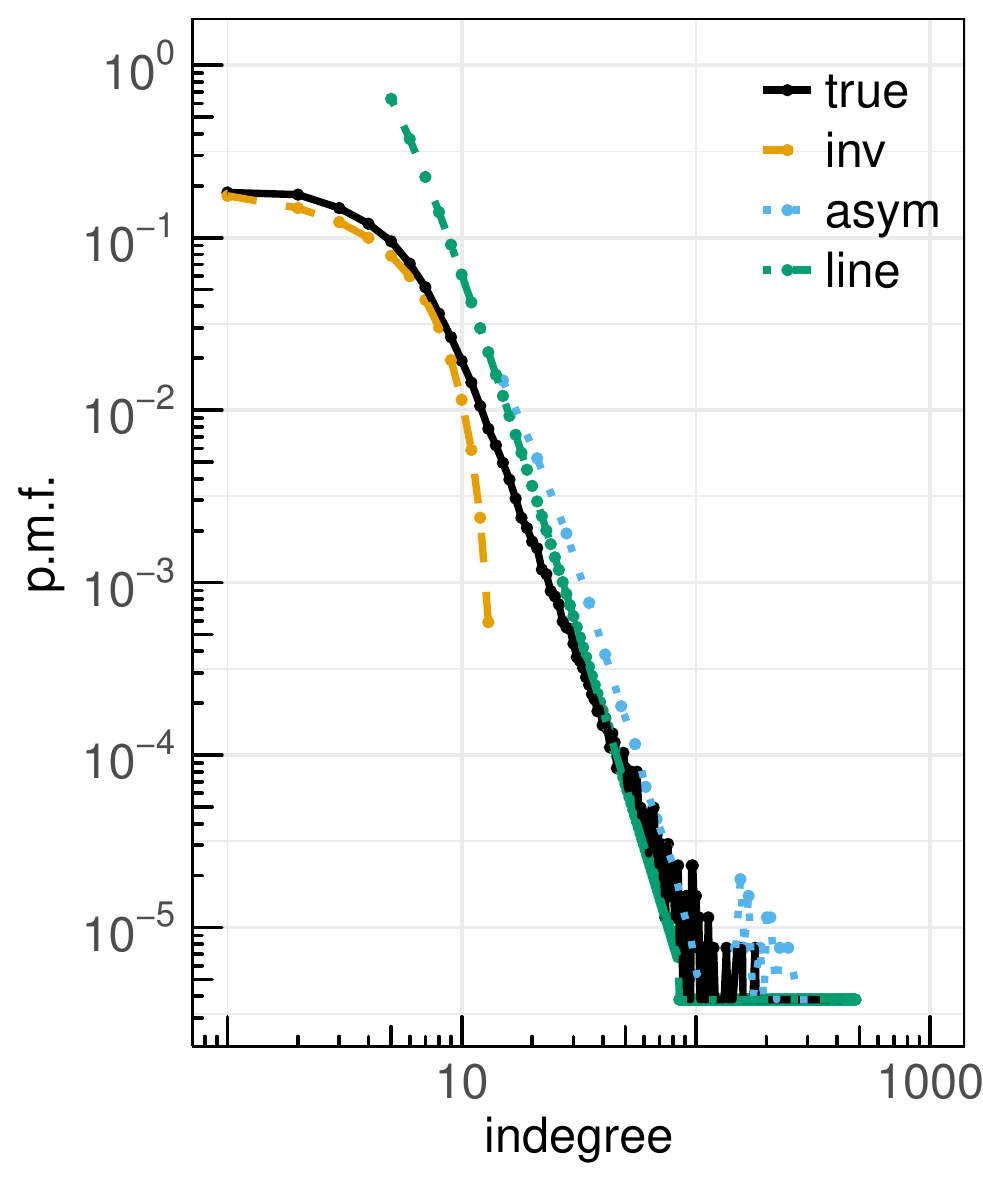}}
  \end{center}
  \caption{Amazon network with RWS ($p=0.15$).}
  \label{f:Amazon}
\end{figure}

Figures \ref{f:RES-WR}--\ref{f:RWS3}  present the (average) estimation of the in-degree distribution for RES-WR, RWS2 and RWS3, respectively, 
where edges are sampled uniformly (in the limit for the RWs).
The inversion approach shows the same behavior as in Figure   \ref{fig:RVS}, with more bias for the smallest in-degrees in the case of the RWS.
We note that the analysis developed in Section \ref{s:asymptotic} for the LINE and ASYM methods is based on  rescaling the tail of the sample in-degree
distribution. For RWS1, the  tail of the  sample in-degree  distribution tends to be smaller compared with RWS2/3 due to MH algorithm
which avoids high-degree vertices; on the other hand   the  tail of the  sample  distribution tends to be smaller with RWS2 than with RWS3 because
of  the jumps of the former  to random edges instead of vertices  which somehow produce a more uniform sample (this explains the differences in   Figures \ref{f:RWS1} ,  \ref{f:RWS2} and \ref{f:RWS3} for the LINE and ASYM methods).
The ASYM method is accurate for large in-degree values. We also found that  the estimation with the inversion approach for all sampling methods is less sensitive  to changes in the jump rate than when using  the asymptotic approach.

Figure \ref{fig:non-power-law} shows the (average) results of  30 generated random directed graphs with exponential in-degree distribution  for RNS-WR and RWS1, where we decreased the sampling probability and jump rate. The number of vertices and edges generated are the same as in the power-law networks above. Only the inversion and ASYM method are used to estimate the in-degree distribution. The inversion approach  shows  better performance over the power-law networks above   due to the form of the p.m.f. at the beginning (decreasing slowly with the  in-degree). A better fit is also shown with  the ASYM method for the tail.
We omit the plots for RES-WR  and RWS2/3 but similar conclusions can be drawn.

\begin{figure}[t]
  \begin{center}
      \subfigure[RWS1 (jump rate: 30\%)]{\label{f:RWS1-cit}\includegraphics[scale=0.51]{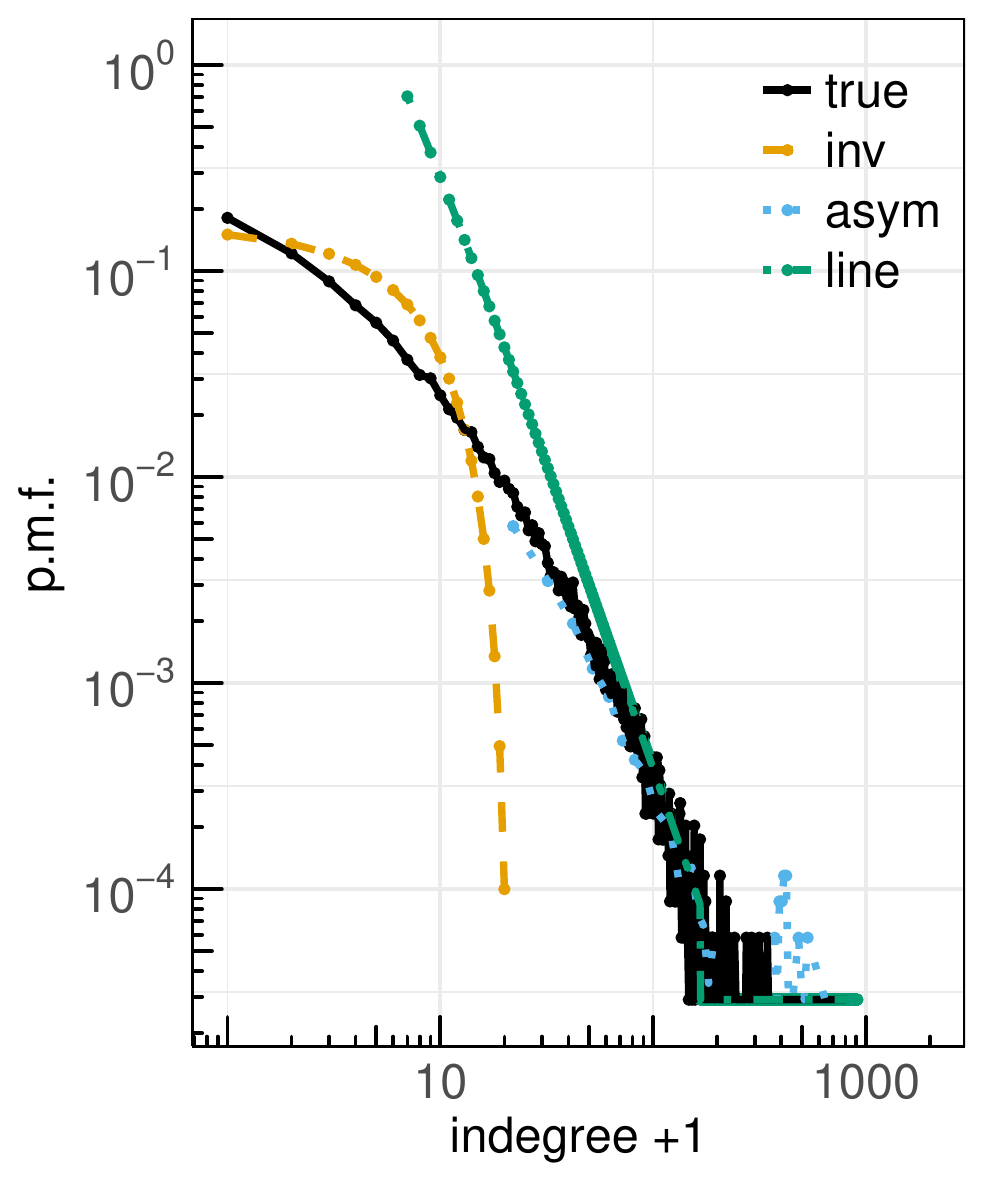}}
      \subfigure[RWS2 (jump rate: 30\%)  ]{\label{f:RWS2-cit}\includegraphics[scale=0.51]{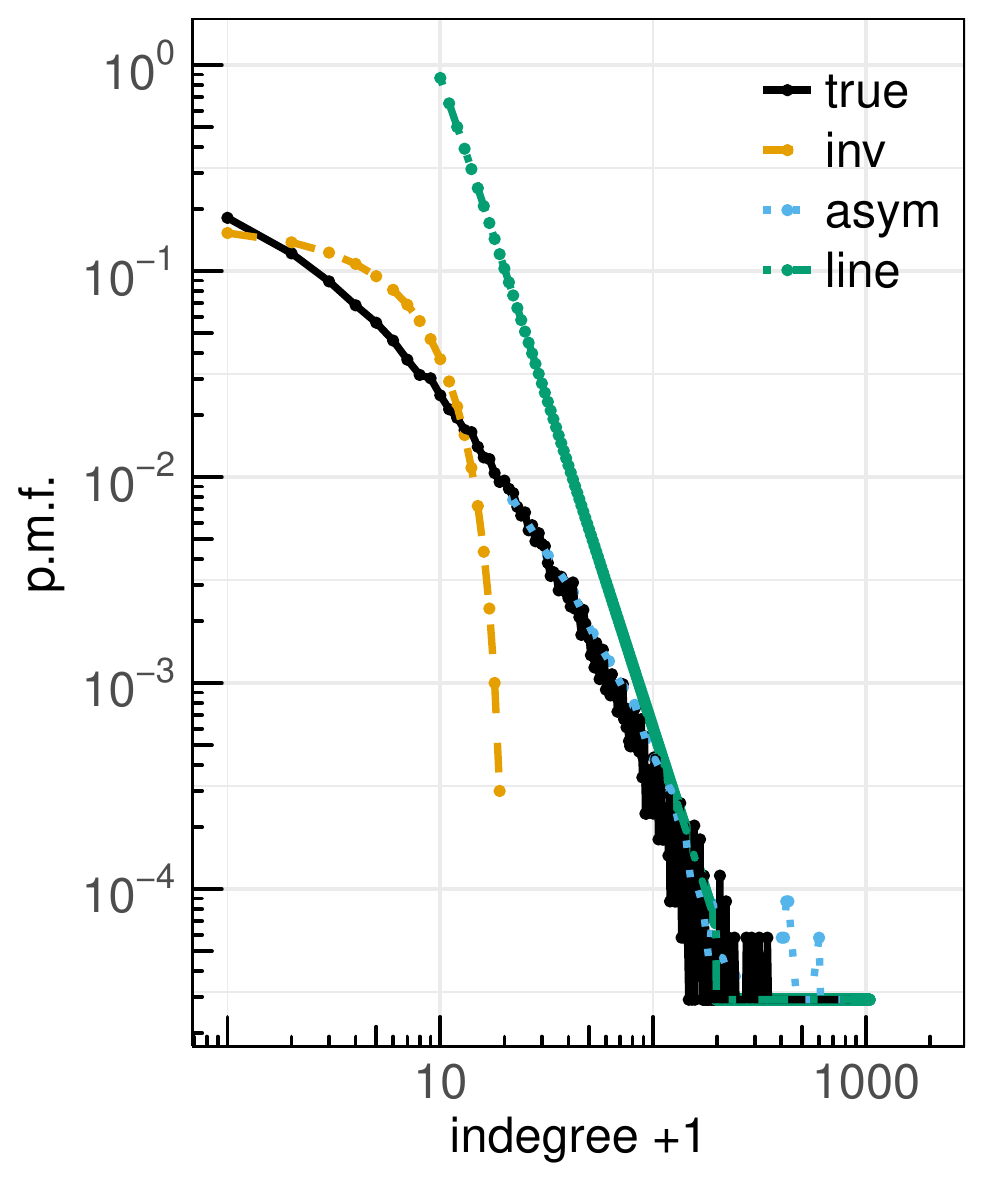}}
  \end{center}
  \caption{ArXiv HEP-PH  network  $(p=0.1)$.}
\end{figure}

\subsection{Real-world directed networks}
\label{s:RealWorldDirectGraphs}
We examine here the suggested estimation methods on two real directed networks. The first network is the Amazon product co-purchasing network discussed in Section \ref{s:intro}.
Figures \ref{f:RWS1-real}-\ref{f:RWS3-real} depict the performance of estimation using our RWS schemes on this network with $p=0.15$ and jump rate 30\%.  Due to the particular  form of the bulk of the in-degree distribution, the inversion method works well for the beginning of the distribution range.  For the LINE method, the parameter $\alpha$ in  (\ref{e:indegree-power-law-estimators-relation-non-log}) is  estimated from the sample in-degree distribution (and the same for the HEP-PH network below). The discussion concerning the power-law networks in Section \ref{s:simulation} above and more specifically on  the differences of the RWS schemes for the asymptotic approach applies here as well.  For  reference and for  comparison   to RWS1 in Figure \ref{f:RWS1-real}, we also include in Figure \ref{f:RVS-real} the estimation results for  RWS-WR. When combined together, the inversion and asymptotic approaches  approaches estimate the whole underlying in-degree distribution.

The second network is  a citation network known as HEP-PH (high energy physics phenomenology)  from the e-print arXiv originally  described in \cite{Gehrke:2003}. The  dataset   covers all the citations in the area for  a  period of ten years with  $N_v =$ 34,546 papers, $N_e =$ 421,578 edges and the mean in(out)-degree of 12.3. If a paper $i$ cites paper $j$, the graph contains a directed edge from $i$ to $j$. If a paper cites, or is cited by, a paper outside the dataset, the graph does not contain any information about this. Figures \ref{f:RWS1-cit}--\ref{f:RWS2-cit} show the estimation based on RWS1 and RWS2, respectively, with $p=0.1$ and jump rate 30\%.
Both sampling schemes  overestimate  the bulk of the distribution with the inversion approach because of the seemingly fast decay of the true distribution and the small sampling probability (not to be recovered from the  sample in-degree distribution which tend to be relatively large for lower in-degrees). The tail of the distribution does not follow that closely a straight line, and therefore, the LINE method only approximates the very end of the tail. On the other hand, due to the large in-degrees, the ASYM method shows a very good  fit to the  tail with the performance  being superior for  RWS1, similarly to what was  observed for the synthetic and Amazon networks considered above. The plots for RVS-WR and RES-WR are omitted since they are similar to Figures 
 \ref{f:RWS1-cit} and \ref{f:RWS2-cit}, respectively.

\section{Discussion}
\label{s:conclusions}

The  estimation of the in-degree distribution of a directed network under  sampling  was investigated in this paper. The problem appears to have received less attention (compared with the inference of the out-degree distribution and degree distribution of undirected networks) in part due to the latent nature of node in-degrees: when a node is sampled, its in-degree is not observed but rather only its out-edges are seen, which contribute to the sample in-degrees of its neighboors. We first considered  two simplest  sampling schemes, random  vertex sampling (RVS) and random  edge sampling (RES), where vertices or edges are selected uniformly at random (with and without replacement). We then constructed three more realistic sampling schemes based on  random walks with jumps  which  in the limit sample vertices or edges  uniformly at random (with replacement). In the latter sense,   RVS and RES could be viewed as  constituting the  basis of our analysis.

The  estimation of the in-degree distribution was formulated  as a linear inverse problem, that involved a matrix dependent on the sampling scheme (RVS or RES) and relating the sample in-degree  distribution to the true underlying in-degree distribution of the network. Due to the inherent  difficulty in observing the node in-degree,    the matrices of the sampling schemes  tended to be  ill-conditioned for small sampling rates.  To deal with ill-conditioning, a penalized weighted least-squares estimation was proposed.
An asymptotic approach was also developed to estimate directly the tail of the  in-degree distribution. This analysis relied on a probabilistic asymptotic equivalence between the true  and sampled in-degree distribution tails. It is much less computationally intensive than using inversion.  Two asymptotic methods  were presented: the LINE method which assumed a power-law behavior found in many real networks, and the ASYM  method which was distribution free.

Finally, our simulations on synthetic networks of different  topologies and applications to real networks showed that the inversion combined with the asymptotic approach can recovered  the true in-degree distribution over its full range under  sampling rates in the range 15\%--20\%. We found that the inversion approach was less sensitive to 
the RW sampling scheme and jump rate used. On power-law networks, the random walk which sampled nodes uniformly (RWS1) performed better  in the estimation of the tail with the asymptotic approach, where  jump rates of 30\% seemed to work well in practice. The jump rates  could be smaller  in the case of  non power-law networks.


As future problems related to this work, for example, it would be interesting to adapt the suggested methodology for quantities of interest in networks other than the in-degree distribution; to get a better sense of how the suggested methodology depends on network characteristics; or to use graph sketching (a sketch is a compact representation of data) in conjunction with sampling to infer even the same in-degree distribution.





\section*{Acknowledgment}

This work was initiated during a sabbatical stay of NA at the Department of Statistics and Operations Research, The University of North Carolina, Chapel Hill. NA is thankful for the hospitality and also acknowledges support by FCT grant CRM:0022222.
SB was partially supported by NSF grants DMS-1613072, DMS-1606839 and ARO grant W911NF-17-1-0010. 
VP was partially supported by NSF grant DMS-1712966.

\begin{appendix}

\section{Proof of the inverse matrix in Eq. (\ref{e:P-NR-RNS-inverse})}
\label{s:appendix}
We will throughout assume that the maximal in-degree $J < n_v$, the number of samples. Further to simplify notation, we write $N:=N_v$ and $n:=n_v$. Recall the expressions of the original matrix $P_s(\cdot,\cdot)$ from \eqref{e:P-NR-RNS} and the corresponding asserted inverse $P_s^{-1}$ from \eqref{e:P-NR-RNS-inverse}. The assertion follows from the following two observations:

{\bf I. Diagonal entries of the product:} For all $0\leq k\leq J$ one has $P_s(k,\cdot) P_s^{-1}(\cdot, k) =1$.

\emph{Proof:} Using the expressions for the two matrices one has 
\[P_s(k,\cdot) P_s^{-1}(\cdot, k)= \frac{\binom{k}{k} \binom{N-k}{n-k}} { \binom{N}{n}} \times \frac{\binom{N-n-1}{0} \binom{N}{k}} { \binom{n}{k}} =1. \]

{\bf II. Off diagonal entries of the product:} For all $k_1 \neq k_2$
\[P_s(k_1,\cdot) P_s^{-1}(\cdot, k_2) =0.\]

\emph{Proof:} Since the matrices are lower triangular, it is enough to show this for $k_1 >  k_2$. Write $k_1 = k_2+A$ for some $A>0$. After some elementary algebraic manipulations, it can be checked that 
\begin{equation}
\label{eqn:binom-expr}
	P_s(k_2+A,\cdot) P_s^{-1}(\cdot, k_2) = c(k_1, k_2) \sum_{k=k_2}^{k_2+A} (-1)^k \binom{M+k_2-1-k}{A-1} \binom{A}{k-k_2},
\end{equation} 
where $c(k_1,k_2)$ is a constant depending on $k_1, k_2$ and $M=N-n$. Thus writing $k^\prime =k-k_2$ and $M^\prime = M-1$, it is enough to show that for all $M^\prime> 0$ and $1\leq A \leq M^\prime$,
\begin{equation}
\label{eqn:to-show}
	\sum_{k^\prime=0}^A (-1)^{k^{\prime}} \binom{M^\prime -k^\prime}{A-1} \binom{A}{k^\prime}= 0.
\end{equation}
Note that the expression on the left of \eqref{eqn:to-show} is of the form,
\[\sum_{k^\prime =0}^{A} (-1)^{k^\prime} P(k^\prime)\binom{A}{k^\prime},\]
where 
\[P(x)= \frac{(M'-x)(M'-x-1)\cdots (M'-x-(A-1))}{(A-1)!},\]
is a polynomial of degree $A-1$ and in particular a polynomial of degree $<A$. Standard results regarding identities of binomial coefficients \cite{ruiz:1996} now imply \eqref{eqn:to-show} and thus \eqref{eqn:binom-expr}. This completes the proof.

\end{appendix}

\bibliographystyle{apalike}

\end{document}